\documentclass[
 reprint,
 superscriptaddress,
 amsmath,amssymb,
 aps,
]{revtex4-1}

\usepackage{graphicx}
\usepackage{dcolumn}
\usepackage{bm}
\usepackage[separate-uncertainty=true]{siunitx}
\usepackage{mhchem}

\begin{document}

\preprint{APS/123-QED}

\title{A general approach to maximise information density in neutron reflectometry analysis}

\author{Andrew R. McCluskey}
  \email{andrew.mccluskey@diamond.ac.uk}
  \email{a.r.mccluskey@bath.ac.uk}
  \affiliation{Diamond Light Source, Rutherford Appleton Laboratory, Harwell Science and Innovation Campus, Didcot, OX11 0DE, UK}
  \affiliation{Department of Chemistry, University of Bath, Claverton Down, Bath BA2 7AY, UK}

\author{Joshaniel F. K. Cooper}
  \affiliation{ISIS Pulsed Neutron and Muon Source, Science and Technology Facilities Council, Rutherford Appleton Laboratory, Harwell Science and Innovation Campus, Didcot, Oxfordshire, OX11 0QX, UK}
 
\author{Tom Arnold}
  \affiliation{European Spallation Source, SE-211 00 Lund, Sweden}
  \affiliation{Department of Chemistry, University of Bath, Claverton Down, Bath BA2 7AY, UK}
  \affiliation{ISIS Pulsed Neutron and Muon Source, Science and Technology Facilities Council, Rutherford Appleton Laboratory, Harwell Science and Innovation Campus, Didcot, Oxfordshire, OX11 0QX, UK}

\author{Tim Snow}
  \affiliation{Diamond Light Source, Rutherford Appleton Laboratory, Harwell Science and Innovation Campus, Didcot, OX11 0DE, UK}
  \affiliation{School of Chemistry, University of Bristol, Bristol, BS8 1TS, UK}

\date{\today}

\begin{abstract}
Neutron and X-ray reflectometry are powerful techniques facilitating the study of the structure of interfacial materials.
The analysis of these techniques is ill-posed in nature requiring the application of model-dependent methods.
This can lead to the over- and under- analysis of experimental data when too many or too few parameters are allowed to vary in the model. 
In this work, we outline a robust and generic framework for the determination of the set of free parameters that are capable of maximising the in-formation density of the model. 
This framework involves the determination of the Bayesian evidence for each permutation of free parameters; and is applied to a simple phospholipid monolayer. 
We believe this framework should become an important component in reflectometry data analysis, and hope others more regularly consider the relative evidence for their analytical models.
\end{abstract}

\maketitle

\section{\label{intro} Introduction}

Neutron and x-ray reflectometry methods offer an important insight into the structure of interfacial materials; such as surfactant systems \cite{mccluskey_bayesian_2019}, biological membranes \cite{clifton_self_2019}, and polymeric photovoltaics \cite{perez_determination_2019}. 
However, the analysis of reflectometry data is a non-trivial task, due in part to the phase problem. 
This is where a direct inversion of the reflectometry profile, to recover the scattering length density, $\rho(z)$ profile is not possible since the phase of the reflected radiation is lost \cite{majkrzak_exact_1995}.
While there has been much work to retrieve the phase information from a reflectometry profile \cite{kirby_phase_2012}; from using reference layers \cite{majkrzak_exact_1995,haan_retrieval_1995,majkrzak_phase_2003,nikova_novel_2019}, polarized neutrons \cite{leeb_determination_1998}, a dwell time approach \cite{fiedeldey_proposal_1992}, and from the investigation of multiple contrasts \cite{majkrzak_exact_1998,majkrzak_first_2000,majkrzak_phase_2003,koutsioubas_model_2019}, these methods are non-trivial and often require a specific experimental setup. 

Where it is not possible to retrieve the phases, a model-dependent analysis method can be used to determine the structure of the experimental system \cite{pedersen_analysis_1994,haan_genetic_1994,nelson_motofit_2006,lee_comparison_2007,gerelli_aurore_2016,gerelli_aurore_2016b,nelson_refnx_2019}.
This is where a model, typically a series of contiguous layers, is created and the model reflectometry is determined using the Abel\`{e}s matrix formalism for stratified media \cite{abeles_propagation_1948} or Parratt recursive method \cite{parratt_surface_1954}.
However, the use of model-dependent methods is ill-posed in nature, where the solution may not necessarily be unique \cite{klibanov_phaseless_1992,klibanov_use_1994}.
This necessitates the integration of \emph{a priori} information about the underlying physics and chemistry of the experimental system, such as assumptions about molecular volume \cite{waldie_localization_2018,campbell_structure_2018,mccluskey_bayesian_2019}. 
This is typically achieved through the re-parameterisation of the model, to describe the system in terms of physiochemical descriptors from which a set of layers can be determined \cite{schalke_structural_2000,heinrich_zooming_2014,heinrich_deuteration_2016}. 
The inclusion of this prior information acts to reduce the dimensionality of the optimisation space and therefore reduces the number of possible structures that would offer satisfactory agreement with the data. 

The inclusion of prior information is often achieved by taking literature values from other techniques and applying them inflexibly to constrain values in the re-parameterised model, such as molecular volumes from computational simulation or solution scattering \cite{sun_order_1994,armen_phospholipid_1998}. 
However, as was shown in the work of Campbell \emph{et al.} \cite{campbell_structure_2018} and McCluskey \emph{et al.} \cite{mccluskey_bayesian_2019}, these values are not always strictly representative when the physical or chemical environment is varied.
This has lead to the use of Markov-chain Monte Carlo (MCMC) methods for the analysis of reflectometry data, resulting in the inclusion of these methods in many common reflectometry analysis packages \cite{hughes_rascal_2017,kienzle_ncnr_2017,nelson_refnx_2019}.
The inclusion of these methods enables the prior probabilities for a series of variables to be outlined, which are considered in a Bayesian fashion during the sampling of the parameter space \cite{bayes_essay_1763,mccluskey_bayesian_2019}. 
This allows for more rational inclusion of the prior information about the experimental system in the analysis, in addition to, the ability to obtain information about the sampled posterior probabilities that are allowed from the experimental data. 

The ability to include ranges of values, in the form of uninformative uniform prior probabilities, or normally distributed priors centred on a literature value in the analysis of an experimental dataset is an important development. 
However, ``with great power, there must also come -- great responsibility!'' \cite{lee_introducting_1962} and the ability to sample or fit all of the parameters in an experimental analysis can lead to a high risk of over-fitting \cite{mayer_drawing_2010}. 
As a result, it is necessary to investigate methods to quantify the information that is available from a given experimental dataset and to enable the comparison of different analytical models; including the number of free parameters that may be considered. 
A specific approach to this problem is outlined in early works from Sivia and co-workers \cite{sivia_analysis_1991,geoghegan_experimental_1996,sivia_bayesian_1998}, where a two-parameter model was compared with a three-parameter model; with the additional parameter being related to the scaling of the model reflectometry. 
Sivia and co-workers assumed that the likelihood function for both of these model would have ``only one significant maximum'', allowing numerical simplifications to be performed and therefore the problem to be solved analytically. 
This may not be possible for all models investigated by reflectometry, due again to the ill-posed nature of the analysis problem. 
Furthermore, the work of Hughes and co-workers \cite{hughes_physical_2019} more recently applied the use of Bayesian model comparison to determine the mixing of lipids in a model bacterial membrane.

In this work, we outline a general Bayesian framework for the comparison of different analytical models for the rationalisation of reflectometry data. 
This method can allow the maximum information to be obtained from the analysis of a given dataset, without risking over-fitting.
In the supplementary information for this paper, we provide detailed code showing how this framework may be applied to a problem leveraging exclusively open-source software packages. 

\section{\label{methods} Methods}

\subsection{\label{bayes_mod} Bayesian model selection framework}

Bayesian model selection involves the comparison of the posterior probabilities $p(H|\mathbf{D})$ \footnote{In this work, a reduced notation has been used, where the background information parameter, $I$ is implied.}, for two models, $H_x$ and $H_y$, given some dataset, $\bm{D}$. 
The posterior probability for a given model, $H$, can be determined from Bayes theorem \cite{bayes_essay_1763}, 
\begin{equation}
    p(H_x|\bm{D}) = \frac{p(\bm{D}|H_x)p(H_x)}{p(\bm{D})},
\end{equation}
where, $p(\bm{D}|H)$ is the evidence for the model given the data, $p(H)$ is the prior probability for the model and $p(\bm{D})$ is the probability associated with the measured data, which is equal to the sum of the evidences for every potential model (which is potentially large if not infinite), 
\begin{equation}
    p(\bm{D}) = \sum_{x = 1}{p(\bm{D}|H_x)}.
\end{equation}
Therefore, in order to describe the evidence as a truely conditional probability, it is necessary to evaluate very possible model that maybe applied to the experimental data. 
This is not feasible, therefore in this work we will use the notation, $p(\bm{D}|H_x)$ for the evidence of a given model, however this is not a strict conditional probability. 
When comparing the models, concerning the same data, the following relationship can be applied \cite{pullen_bayesian_2014}, 
\begin{equation}
    \frac{p(H_x|\bm{D})}{p(H_y|\bm{D})} = \frac{p(\bm{D}|H_x)}{p(\bm{D}|H_y)} \times \frac{p(H_x)}{p(H_y)},
    \label{equ:ratio}
\end{equation}
where, the probabilities associated with the data only cancel out, and the relative posterior probabilities depend only on the evidence and prior probabilities for each model. 
Throughout this work, we have used the uninformative assumption that the prior probability for all models is \num{1}.
Therefore, it is only the evidence for each model, for the given data, that influences the posterior probability. 

Nested sampling is a method for Bayesian evidence determination that was developed by Skilling \cite{skilling_nested_2006} and is implemented in the Python package \texttt{dynesty} \cite{speagle_dynesty_2019}.
This sampling method allows for the evidence determination through the estimation of the, possibly multi-dimensional, integral under the likelihood, $\mathcal{L}(\bm{X}|H)$, and parameter specific prior $p(\bm{X}|H)$ \cite{sivia_data_2005}, 
\begin{equation}
    p(\mathbf{D}|H) = \iint_{\bm{R}}{\mathcal{L}(\bm{X}|H)p(\bm{X}|H)}\;\text{d}^M\bm{X},
\end{equation}
where, $\bm{X}$ is a vector of length $M$ of the varying parameters in the model, $\bm{R}$ is a $2 \times M$ matrix that describes the range over which the integral should be evaluated and $p(\bm{X}|H)$ is the parameter specific prior. 
Nested sampling involves ``intelligently'' sampling this integral until some stopping criterion is achieved.
This stopping criterion is when the following is less than a given value,
\begin{equation}
    \begin{aligned}
        \Delta \ln\{\hat{p}(\bm{D}|H)_i\} = \ln\{\hat{p}(\bm{D}|H)_i - \Delta \hat{p}(\bm{D}|H)_i\} \\
        - \ln \hat{p}(\bm{D}|H)_i, 
    \end{aligned}
\end{equation}
where, $\hat{p}(\bm{D}|H)_i$ is the current estimated evidence and $\Delta \hat{p}(\bm{D}|H)_i$ can be estimated from the prior volume of the previous point in the sampling. 
In this work, a stopping criteria of $\Delta \ln\{\hat{p}(\bm{D}|H)\} = 0.5$ was used. 

The likelihood is the difference between the data and model, and is defined in this work as it is implemented in the \texttt{refnx} reflectometry package \cite{nelson_refnx_2019,refnx_0.1.7_2019},
\begin{equation}
    \begin{aligned}
        \ln\mathcal{L} = -\frac{1}{2}\sum_{i=1}^{N}\Bigg(\bigg\{\frac{R(q_i) - R(q_i)_m}{\delta R(q_i)}\bigg\}^2 \\ + \ln[2\pi\{\delta R(q_i)\}^2]\Bigg),
    \end{aligned}
\end{equation}
where, $N$ is the number of measured $q$-vectors, $q_i$ is the $i$th $q$-vector, $R(q_i)$ is the experimental reflected intensity at the $i$th $q$-vector, $\delta R(q_i)$ is the uncertainty in the experimental reflected intensity at the $i$th $q$-vector, and $R(q_i)_m$ is the model reflected intensity at the $i$th $q$-vector determined from the implementation of the Abel\`{e}s matrix formalism for stratified media \cite{abeles_propagation_1948,parratt_surface_1954} in \texttt{refnx}.

The evidence allows for the comparison of two models, where the data are the same.
It is only possible to compare a given pair of models for the same data, as in Equation~\ref{equ:ratio} the data-only dependent probabilities were cancelled out as they were equal. 
The Bayes factor, $B_{xy}$, is the name given to the ratio between the evidence for two models, 
\begin{equation}
    \begin{aligned}
        \ln{B_{xy}} & = \ln{\Bigg[\frac{p(\bm{D}|H_x)}{p(\bm{D}|H_y)}\Bigg]} \\
        & = \ln[p(\bm{D}|H_x)] - \ln[p(\bm{D}|H_y)]. 
    \end{aligned}
\end{equation}
The interpretation of the Bayes factor was outlined by Kass and Raftery \cite{kass_bayes_1995}, which is given in Table~\ref{tab:kass}.
\begin{table}
\caption{\label{tab:kass} Interpretation of the Bayes factor, $B_{xy}$ between models $x$ and $y$.}
\begin{ruledtabular}
\begin{tabular}{lr}
\textrm{$2\ln{B_{xy}}$} & Evidence against $H_y$\\
\colrule
$[0, 2)$ & Not worth more than a bare mention \\
$[2, 6)$ & Positive \\
$[6, 10)$ & Strong \\
$[10, \infty)$ & Very strong \\
\end{tabular}
\end{ruledtabular}
\end{table}
This method is easily generalised to any model-dependent experiment, and has been applied in areas such as systems biology, astronomy, computational chemistry, and indeed, in a less general fashion than the current work, to neutron reflectometry \cite{pullen_bayesian_2014,cornish_tests_2007,ensign_bayesian_2010,sivia_bayesian_1998}. 

The Bayesian approach allows for the comparison of models where there are different numbers of free parameters, without the risk of overfitting and ``fitting an elephant'' \cite{mayer_drawing_2010}.
This is due to the Bayesian evidence being derived from an integral in parameter space, and therefore scaling with the number of parameters. 
This means that the addition of an additional free parameter to the model must offer a significant improvement to the likelihood determined.
However, it is important also to note that the accuracy of the determined evidence depends on the prior probabilities chosen for each of the free parameters and therefore care should be taken to ensure that these are meaningful. 

\subsection{\label{nr_meas} Neutron reflectometry measurements}

The neutron reflectometry measurements investigated in this work were previously published by Hollinshead \emph{et al.} \cite{hollinshead_effects_2009} and subsequently re-analysed by McCluskey \emph{et al.} \cite{mccluskey_assessing_2019}. 
The measurements concern the study of a monolayer of 1,2-distearoyl-\emph{sn}-glycero-3-phosphocholine (DSPC) at the air/water interface and were conducted on seven different isotopic contrasts of the phospholipid and water. 
These contrasts were made up from four phospholipid types; fully-hydrogenated (\ce{h}-DSPC), head-deuterated (\ce{d_{13}}-DSPC), tail-deuterated (\ce{d_{70}}-DSPC), and fully deuterated (\ce{d_{83}}-DSPC).
These were paired with two water contrasts; \ce{D2O} and air-contrast matched water (ACMW, where \ce{D2O} and \ce{H2O} are mixed such that the resulting scattering length density is zero).
The pairing of the fully-hydrogenated phospholipid with ACMW was not performed, most likely due to the lack of the scattering available from such a system.
Table~\ref{tab:shorthand} gives the shorthands that are used in this work to refer to the different contrast pairings investigated.
\begin{table}
\caption{\label{tab:shorthand} The shorthands used for the different combinations of phospholipid monolayer and water investigated.}
\begin{ruledtabular}
\begin{tabular}{lcr}
Shorthand & Phospholipid contrast & Water contrast \\
\colrule
\ce{h}-\ce{D2O} & h-DSPC & \ce{D2O} \\
\ce{d_{13}}-\ce{D2O} & \ce{d_{13}}-DSPC & \ce{D2O} \\
\ce{d_{13}}-\ce{ACMW} & \ce{d_{13}}-DSPC & \ce{ACMW} \\
\ce{d_{70}}-\ce{D2O} & \ce{d_{70}}-DSPC & \ce{D2O} \\
\ce{d_{70}}-\ce{ACMW} & \ce{d_{70}}-DSPC & \ce{ACMW} \\
\ce{d_{83}}-\ce{D2O} & \ce{d_{83}}-DSPC & \ce{D2O} \\
\ce{d_{83}}-\ce{ACMW} & \ce{d_{83}}-DSPC & \ce{ACMW} \\
\end{tabular}
\end{ruledtabular}
\end{table}
The data analysed in this work was taken at a surface pressure of \SI{30}{\milli\newton\meter^{-1}}.
Additional details of the data collection may be found in the original publication \cite{hollinshead_effects_2009}.

\subsection{\label{models} Analytical model}

In the model-dependent analysis of neutron reflectometry data, it is necessary to have a description of the experimental system to which the Abel\`{e}s matrix formalism may be applied. 
Campbell \emph{et al.} investigated different models for the study of phospholipid monolayers and found a two-layer description that differentiates between the phospholipid heads, $h$, and tails, $t$, to be the best at reproducing the reflected intensity \cite{campbell_structure_2018}. 
The head layer is adjacent to the solvent, while the tail layer is adjacent to the air. 
The thickness of the tail layer is defined as $d_t$ and the head layer is $d_h$.
At each of the three interfaces, there is some interfacial roughness, $\sigma$, which is modelled with an error function \cite{nevot_caracterisation_1980}.
In this work, the roughness was taken to be conformal, which is to say that it does not vary between interfaces, in agreement with the work of Campbell \emph{et al.} \cite{campbell_structure_2018}.
This final assumption is reasonable given the fact that this system consists of a monolayer of a single lipid type, however, this may not be reasonable in more complex systems. 

The scattering length density, $\rho_i$, for a layer $i$, from which the model reflectometry profile is obtained, is calculated from the volume fraction of the chemical component, $\phi_i$, the scattering length of the component, $b_i$ (found in Table~\ref{tab:sl}), and its molecular volume, $V_i$, 
\begin{equation}
       \rho_i = \frac{b_i}{V_i}\phi_i + \rho_{\ce{H2O}} (1 - \phi_i).
       \label{equ:sld}
\end{equation}
where, $\rho_{\ce{H2O}}$ is the scattering length density of water, which is \SI{0}{\angstrom^{-2}} for ACMW and \SI{6.35e-6}{\angstrom^{-2}} for \ce{D2O}.
To preserve the chemical bonding between the heads and tails of the phospholipid, the following constraint is included in the model, 
\begin{equation}
    \phi_h = \frac{d_tV_h\phi_t}{V_td_h}.
    \label{equ:constrain}
\end{equation}
This ensures that the number of head groups is equal to the number of pairs of tail groups. 
\begin{table}
\caption{\label{tab:sl} The scattering lengths, $b_i$, for each of the DSPC phospholipid components in this work; where the subscript indicates either the head, $h$, or tail, $t$.}
\begin{ruledtabular}
\begin{tabular}{lcr}
Phospholipid contrast & $b_h$/\si{\femto\meter} & $b_t$/\si{\femto\meter} \\
\colrule
\ce{h}-DSPC & \num{6.01} & \num{-3.58} \\
\ce{d_{13}}-DSPC & \num{19.54} & \num{-3.58} \\ 
\ce{d_{70}}-DSPC & \num{11.21} & \num{69.32} \\ 
\ce{d_{83}}-DSPC & \num{24.75} & \num{69.32} \\
\end{tabular}
\end{ruledtabular}
\end{table}

The model outlined above has up to six possible free parameters; the two-layer thicknesses, $d_t$ and $d_h$, the two molecular volumes, $V_t$ and $V_h$, the volume fraction of the tail layer, $\phi_t$, and the conformal interfacial roughness, $\sigma$. 
In this work, the evidence as a function of these parameters is of interest as it is possible to use the evidence and Bayes factor to determine the best set of free parameters for a given set of data. 
Therefore, in Table~\ref{tab:ref} the uniform prior probability range and value when constrained is given for each of the parameters. 
\begin{table}
\caption{\label{tab:ref} The constrained and uniform prior probability range for each of the free parameters in the analytical model.}
\begin{ruledtabular}
\begin{tabular}{lcr}
Parameter & Constrained Value & Prior Range \\
\colrule
$d_h$/\si{\angstrom} & \num{10.0}\footnote{The work of Campbell \emph{et al.} uses a head layer thickness of \SI{10}{\angstrom} for the phosphatidylcholine head layer based on ``molecular dimensions'' \cite{campbell_structure_2018}.} & $[8.0, 16.0)$ \\
$V_h$/\si{\angstrom^3} & \num{339.5}\footnote{From the work of Nu\v{c}erka \emph{et al.} \cite{kucerka_determination_2004} and Balgav\'{y} \cite{balgavy_evaluation_2001}.} & $[300.0, 380.0)$ \\
$d_t$/\si{\angstrom} & \num{21.0}\footnote{The maximum tail length for DSPC is found as \SI{24.3}{\angstrom} from the Tanford equation \cite{tanford_hydrophobic_1980}, the value of \SI{21}{\angstrom} comes from assuming a chain tilt of \SI{30}{\degree}.} & $[10.0, 26.0)$ \\
$\phi_t$ & \num{1.0}\footnote{Based on the hydrophobic interaction of the tail groups, it is unlikely that water would intercalate.} & $[0.5, 1.0)$\footnote{A maximum of \num{1} was set as a volume fraction greater than \num{1} would be unphysical.} \\
$V_t$/\si{\angstrom^3} & \num{850.4}\footnote{From the previous analysis of this dataset \cite{mccluskey_assessing_2019}, which agrees with the expected compression of \SI{\sim12}{\percent} discussed by Campbell \emph{et al.} \cite{campbell_structure_2018}.} & $[800.0, 1000.0)$ \\
$\sigma$/\si{\angstrom} & \num{2.9}\footnote{From the work of Campbell \emph{et al.} \cite{campbell_structure_2018}.} & $[2.9, 5.0)$\footnote{The minimum of \SI{2.9}{\angstrom} is based on that of water \cite{braslau_surface_1985,sinha_xray_1988}, and the fact that surfactants have been shown to increase surface roughness \cite{tikhonov_xray_2000}.} \\
\end{tabular}
\end{ruledtabular}
\end{table}
These six parameters can give rise to \num{63} permutations where there is at least one free parameter. 
These permutations constitute every possible combination of the six parameters, either held constant or being allowed to varies within the prior probability. 
This allows for a complete understanding of which combinations of constrained and free parameters give rise to the highest evidence, and therefore represents the maximum information density that may be obtained from the given data.
We note again, that both the chosen prior and the constrained values are not infallible and will influence the result. 

All seven isotopic neutron contrasts were analysed with a single structural model. 
In doing so we make the common assumption that there is no effect on the structure as a result of the contrast variation method \cite{nelson_motofit_2006}. 
Additionally, the reflected intensity from the model was scaled and a uniform background was added based on values obtained in the previous analysis of this dataset \cite{mccluskey_assessing_2019}. 
We accept that these parameters may influence the resulting evidence, as was shown in the work of Sivia and Webster \cite{sivia_bayesian_1998}.

\section{\label{results} Results \& Discussion}

\subsection{All parameters}
\begin{figure}
\includegraphics[width=0.4\textwidth]{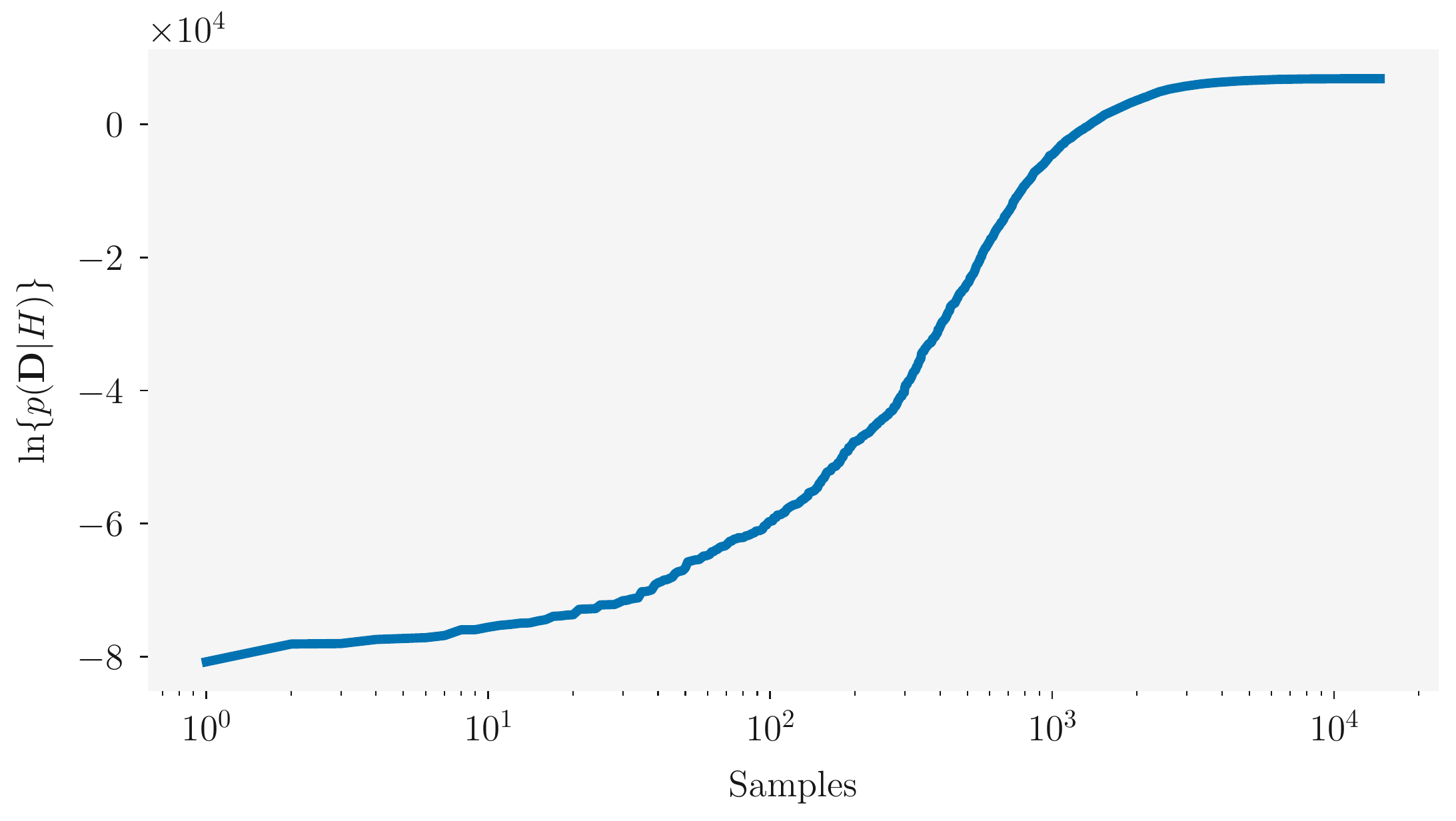}
\caption{\label{fig:iterations} The estimation of evidence as a function of number of samples; for the model where all parameters were unconstrained.}
\end{figure}

Figure~\ref{fig:iterations} shows the variation of the estimated evidence as a function of iterations, for the model where all six parameters were unconstrained.
This plot shows clearly the ability for the nested sampling method to determine a good estimate of the evidence, with the plot plateauing at \num{6859.9\pm0.3}\unskip\;after more than \num{10000} samples.
This was the slowest model to converge, as it was sampling the high-dimensionality parameter space, on a workstation computer this nested sampling process took around \SI{20}{\minute} to converge leading to the results provided here.

\begin{figure}
\includegraphics[width=0.45\textwidth]{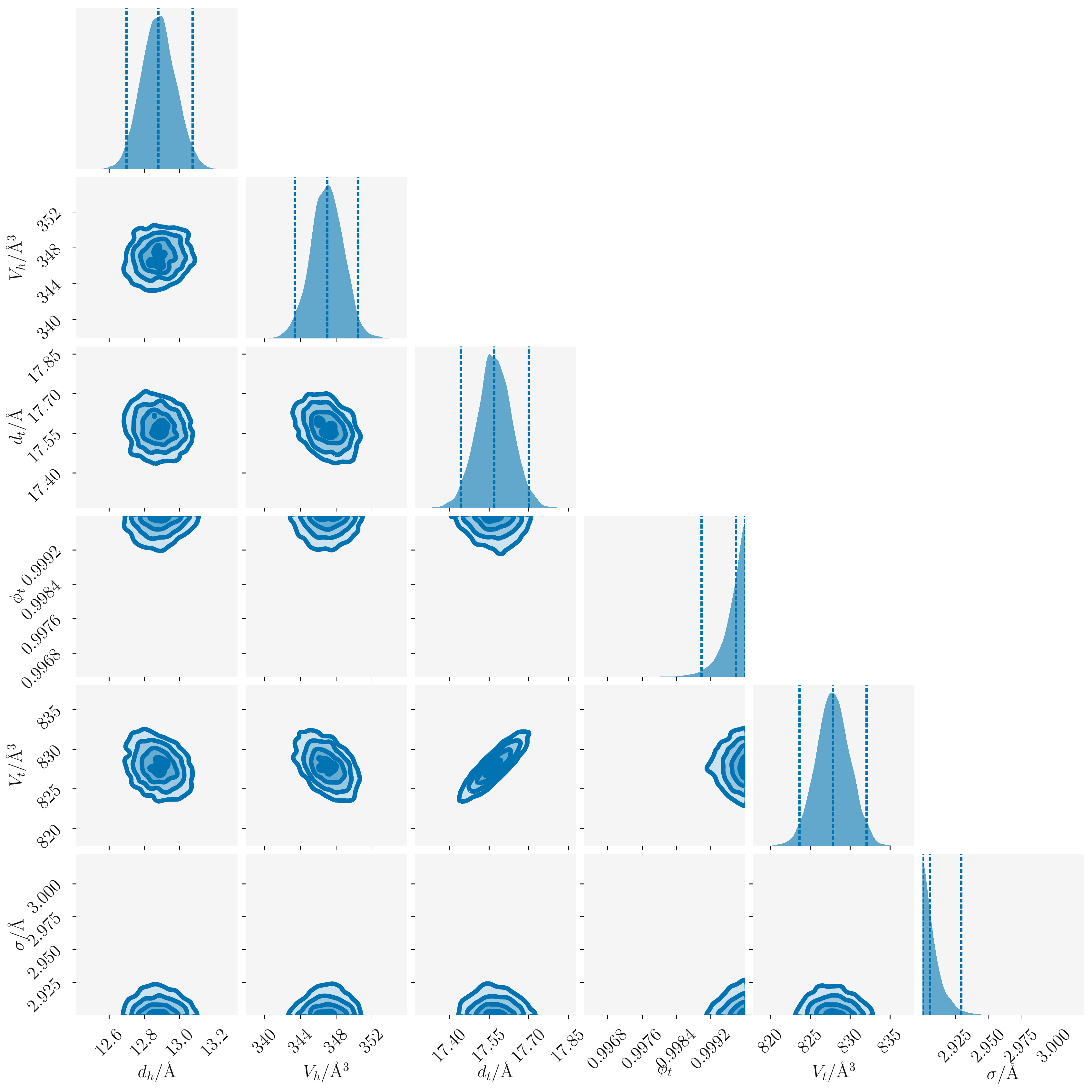}
\caption{\label{fig:post} The posterior probabilities and correlation plots for each of the six variables in the model.}
\end{figure}

In addition to the ability for the nested sampling method to determine the evidence for a given model, it also allows for the posterior probabilities of the analysis to be evaluated. 
Figure~\ref{fig:post} gives the posterior probabilities and cross-correlations between each of the six parameters. 
It is clear that the volume fraction of the tail layer, $\phi_t$, and the interfacial roughness, $\sigma$, both are not complete represented by the defined prior probabilities. 
Hence, there lacks a full Gaussian distribution and the extremely skewed probability distribution. 
However, maximum for $\phi_t$ and minimum for $\sigma$ both come from physical limitations on the system; the former being that the volume fraction cannot be greater than 1, and the latter being that a surfactant monolayer will increase the surface roughness compared to water \cite{tikhonov_xray_2000}, which is \SI{\sim2.9}{\angstrom} \cite{braslau_surface_1985,sinha_xray_1988}. 
The skew-ness of $p(\phi_t)$ is to be expected given the hydrophobic interaction between the phospholipid tails which create the situation where it is highly unlikely for water to be intercalated into the tail structure. 
The nature of $p(\sigma)$ is similar in that the expected value of \SI{2.9}{\angstrom} is the minimum value available to the distribution. 
However, it may be the case that by fitting all possible parameters, the data is being overfitted. 
Luckily, the Bayesian model comparison method allows for the evidence comparison to fit the model which offers the maximum information to be obtained, without over-fitting to the data. 

\subsection{Greatest evidence}
\begin{figure*}
\includegraphics[width=\textwidth]{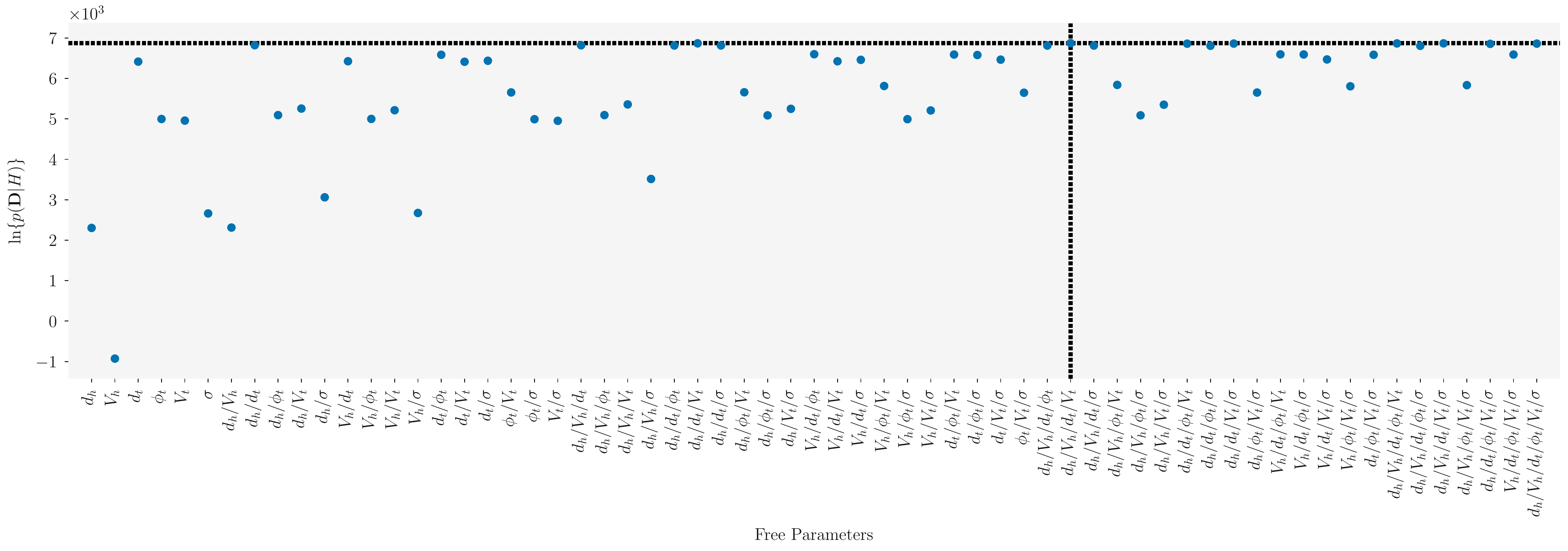}
\caption{\label{fig:evidence} A comparison of the evidence for each of the \num{63} permutations of free parameters available to the analytical model. The dashed line indicated the maximum evidence observed.}
\end{figure*}

The evidence for all of the \num{63} possible permutations of the free parameters was determined.
This allows for the comparison of these to determine which model (set of varying parameters) offers the most information from the dataset, without the over-fitting to the data that may be occurring when all six parameters are varied. 
This comparison plotted in Figure~\ref{fig:evidence} \footnote{These values are tabulated in Table S1.}. 
The dotted black line indicates the set of free parameters that offer the greatest evidence for the co-refinement of the seven contrasts of neutron reflectometry data, which in this case is $d_h$/$V_h$/$d_t$/$V_t$\unskip, which has an evidence of \num{6872.1\pm0.2}\unskip. 
The set of parameters with the next greatest evidence is $d_h$/$d_t$/$V_t$\unskip, however, the evidence for this set of parameters is \num{6866.8\pm0.2}\unskip.
The interpretation of the Bayes factor between these two sets of parameters gives $2\ln{B_{xy}}=$\num{10.7\pm0.7}\unskip\;where, $d_h$/$V_h$/$d_t$/$V_t$\unskip\;is $x$ and $d_h$/$d_t$/$V_t$\unskip\;is $y$. 
This indicates that there strong or very strong evidence for the more complex model, and that the maximum information density that can be obtained from the analysis at hand, without over-fitting, is the fitting of $d_h$/$V_h$/$d_t$/$V_t$\unskip. 

\begin{figure}
\includegraphics[width=0.45\textwidth]{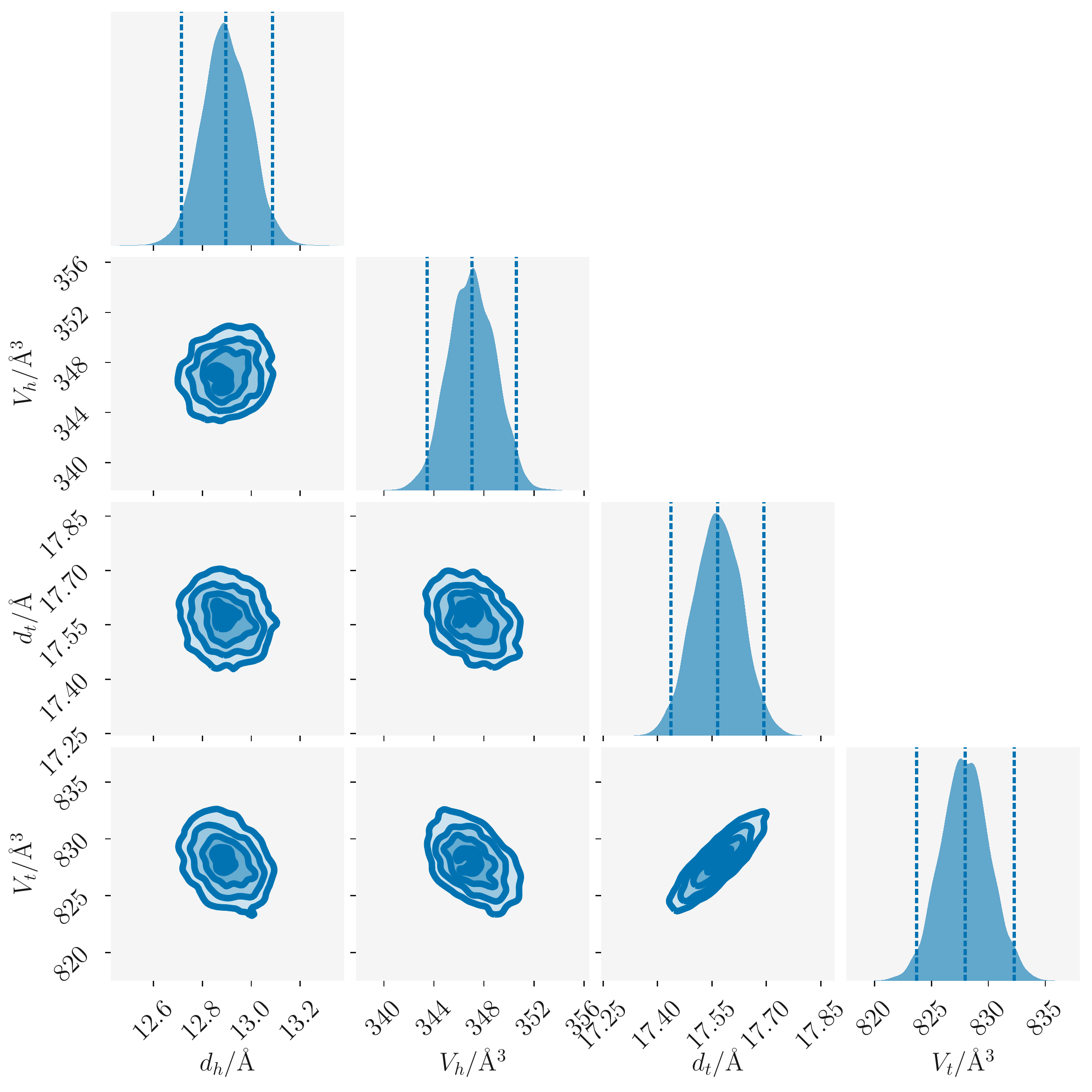}
\caption{\label{fig:post_best} The posterior probabilities and correlation plots when \protect\input{best_label.txt}\unskip\;were the variable parameters.}
\end{figure}

Figure~\ref{fig:post_best} shows the posterior probability distributions that were obtained from the $d_h$/$V_h$/$d_t$/$V_t$\unskip\;set of parameters. 
The median values for each of the probability distributions and their \SI{95}{\percent} confidence intervals are given in Table~\ref{tab:results}. 
\begin{table}
\caption{\label{tab:results} The median values and \SI{95}{\percent} confidence intervals for each of the free parameters in the model with the greatest evidence.}
\begin{ruledtabular}
\begin{tabular}{lr}
Free Parameter & Value \\
\colrule
$d_h$/\si{\angstrom} & $12.9^{+0.2}_{-0.2}$ \\ 
$V_h$/\si{\angstrom^3} & $347.0^{+3.6}_{-3.6}$ \\ 
$d_t$/\si{\angstrom} & $17.6^{+0.1}_{-0.1}$ \\ 
$V_t$/\si{\angstrom^3} & $828.0^{+4.3}_{-4.2}$ \\ 
\end{tabular}
\end{ruledtabular}
\end{table}
The value found here for the phosphatidylcholine is slightly larger than the \SI{10}{\angstrom} used by Campbell \emph{et al.} \cite{campbell_structure_2018} and the \SI{10.5}{\angstrom} quoted by Li \emph{et al.} \cite{li_some_1998}. 
The thickness of the tail layer that is observed is less than that suggested previously \cite{campbell_structure_2018}.
However, it agrees well with the results from the original analysis of this data \cite{hollinshead_effects_2009}, which itself was consistent with previous x-ray reflectometry measurements \cite{dewolf_phase_1999,brezesinski_xray_2001,struth_organization_2001}.
The tail molecular volume agrees well with the compression of \SI{\sim12}{\percent}, when compared to the values from other experimental techniques \cite{sun_order_1994,armen_phospholipid_1998}, suggested by Campbell \emph{et al.}.

\begin{figure}
\includegraphics[width=0.4\textwidth]{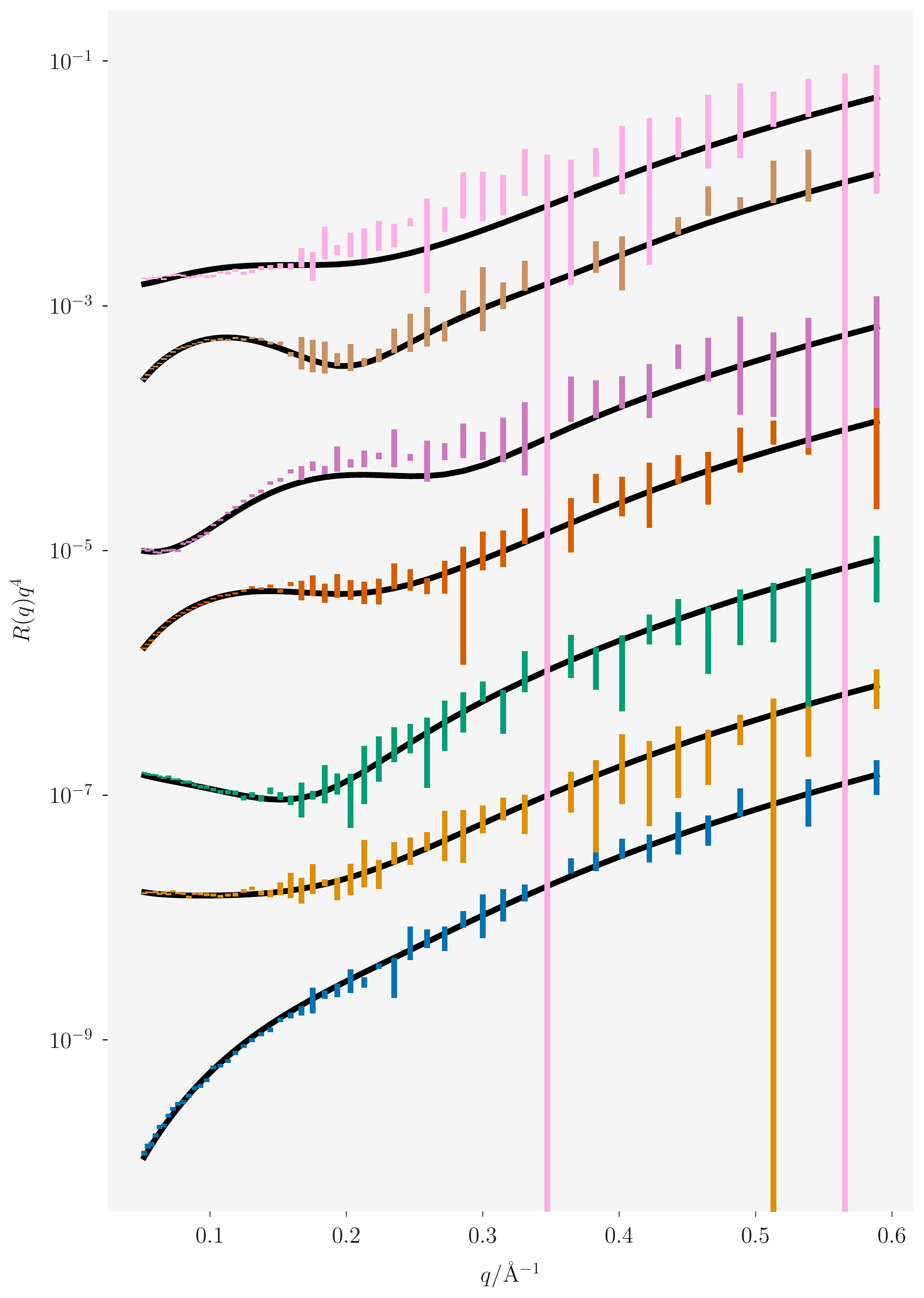}
\caption{\label{fig:refl} The experimental (coloured lines) reflectometry and the median values for the model with greatest evidence (black lines). The different contrasts are offset be an order of magnitude in reflected intensity; \ce{d_{13}}-ACMW (blue), \ce{d_{13}}-\ce{D2O} (orange), \ce{h}-\ce{D2O} (green), \ce{d_{70}}-ACMW (red), \ce{d_{70}}-\ce{D2O} (purple), \ce{d_{83}}-ACMW (brown), and \ce{d_{83}}-\ce{D2O} (pink).}
\end{figure}

The reflectometry profiles for the median values is compared with the experimental data in Figure~\ref{fig:refl}. 
There is clear agreement between the experiment and the model across all of the contrasts. 
This indicates that it is likely that there is little variation due to the isotopic contrast variation that was used, and the assumption that is discussed in \ref{models} is valid. 

\subsection{Comparison with previous analysis}

As mentioned above, this particular experimental dataset was previously analysed by McCluskey \emph{et al.} \cite{mccluskey_assessing_2019}.
The analytical model used in this previous work was similar to that used herein, however, the tail volume of the phospholipid was constrained based on the assumption that the measured area per molecule was accurate and the volume fraction of the tail layer is \num{1}. 
That model had three free parameters, namely $d_h$, $d_t$, and $\sigma$. 

This previous analysis made no effort to evaluate different models of analysis or numbers of free parameters. 
Therefore, it is not clear if the three parameters chosen offered the maximum information density available. 
While not completely comparable, as the constrained values in this work were different to those used previously, the suggestion from this work that having $d_h$/$V_h$/$d_t$/$V_t$\unskip\;as the free parameters do not match with those probed previously.
In future, we hope that by outlining this general methodology, others will consider the evidence of their particular model and the free parameter being used during their analysis.

\subsection{Best per number of parameters}

\begin{figure}
\includegraphics[width=0.4\textwidth]{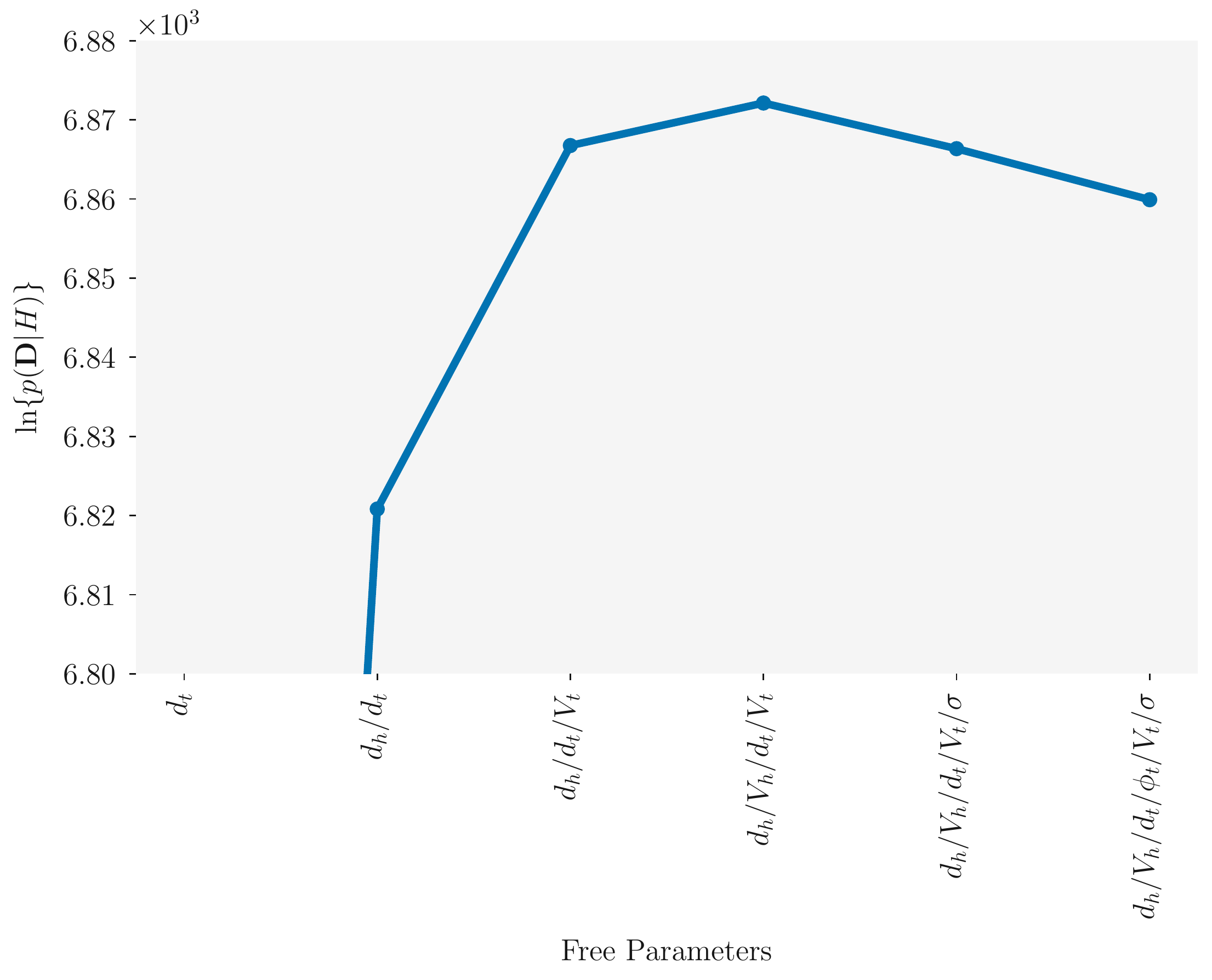}
\caption{\label{fig:best_per} A comparison of the evidence for the set of parameters with the greatest evidence, per number of parameters. The evidence for a single parameter has been cut off for clarity, however, it had a value of \protect\input{tt_ev.txt}.}
\end{figure}

Finally, we will briefly present the best set of parameters for each number of parameters. 
These are shown in Figure~\ref{fig:best_per}, which has the characteristic shape found in the previous work by Sivia and others \cite{sivia_analysis_1991,sivia_introduction_1993,sivia_bayesian_1998,sivia_data_2005}. 
This plot can be seen to peak when there are four free parameters, before dropping off slowly. 
This drop off indicates that when more parameters are included there is an indication that over-fitting is occurring. 
This plot shows clearly where the maximum information is available in the analysis for this particular reflectometry data. 

\section{\label{conclusions} Conclusions}

This work outlines a general application of a Bayesian framework to the problem of model selection in neutron reflectometry. 
Building on the work of Sivia and co-workers in the 1990s \cite{sivia_bayesian_1998}, we have utilised a nested sampling method to generalise the methodology beyond the assumption of only one significant maximum \cite{sivia_analysis_1991}. 
This offers the ability to go beyond the three-parameter space previously investigated to consider every possible parameter in an analysis problem.
In doing so, it is possible to compare the evidence of different models, where different numbers of parameters are allowed to vary, to determine that which offers the maximum information, without over-fitting. 

To show this in action, we have investigated the analysis of seven isotopic contrasts of neutron reflectometry data from a phospholipid monolayer at the air/water interface. 
This particular analysis had a maximum of six variable parameters, and in this work, the evidence of all \num{63} possible permutations of these parameters was determined. 
This allowed for the permutation that offered the maximum evidence to be found, which under the circumstances of the chosen priors was when $d_h$/$V_h$/$d_t$/$V_t$ were varied. 
This particular combination of parameters had an evidence of \num{6872.1\pm0.2}, while the next best, which had more free parameters had a value of \num{6866.8\pm0.2}. 
This was interpreted in terms of the Bayes factor, and it was shown that there was strong evidence for the $d_h$/$V_h$/$d_t$/$V_t$ model.  
Finally, we have compared these free parameters with those used previously to analyse the same dataset \cite{mccluskey_assessing_2019}. 
This comparison indicates that this previous analysis did not necessarily maximise the information density available to the data. 

We accept that the relative simplicity of this model system allows for the use of nested sampling on a relatively short timescale, with all of the computation capable of being performed in a single day on a workstation machine. 
However, the applicability of the methods discussed herein is relevant to any model-dependent reflectometry analysis, but we note that as the number of free parameters increases so does the dimensionality of the parameter space being sampled, leading to longer computational timescales. 
Therefore, we propose that for analyses with a large number of parameters the analyst starts by reducing the number of free parameters as much as is possible to ensure that the nested sampling will converge in a reasonable timescale and gradually increasing the complexity. 
We hope that by developing and sharing of this methodology, in future those analysing reflectometry data will consider which of the free parameters in their models may be varied and will try to determine the evidence for their particular model for other options before completing their analysis. 

\section*{Reproducibility Statement}

Electronic Supplementary Information (ESI) available: All analysis/plotting scripts and data files allowing for a fully reproducible, and automated, analysis workflow for the work presented is available at \url{https://github.com/arm61/model_select} (DOI: 10.5281/zenodo.3820690) under a CC BY-SA 4.0 license.

\section*{Author Contributions}

A.R.M. proposed, implemented, and applied the Bayesian model comparison framework, with input from T.A., J.F.K.C. and T.S.; J.F.K.C. and T.S. sourced funding; A.R.M. wrote the manuscript, with input from all authors.

\section*{Acknowledgements}

A.R.M. would like to acknowledge David J. Barlow and M. Jayne Lawrence for kindly sharing the neutron reflectometry data used, and Simon Titmuss for suggesting the use of the \texttt{dynesty} package.
This work is supported by the Ada Lovelace Centre – a joint initiative between the Science and Technology Facilities Council (as part of UK Research and Innovation), Diamond Light Source, and the UK Atomic Energy Authority.

\bibliography{paper}

%merlin.mbs apsrev4-1.bst 2010-07-25 4.21a (PWD, AO, DPC) hacked
%Control: key (0)
%Control: author (8) initials jnrlst
%Control: editor formatted (1) identically to author
%Control: production of article title (-1) disabled
%Control: page (0) single
%Control: year (1) truncated
%Control: production of eprint (0) enabled
\providecommand{\noopsort}[1]{}\providecommand{\singleletter}[1]{#1}%
\begin{thebibliography}{64}%
\makeatletter
\providecommand \@ifxundefined [1]{%
 \@ifx{#1\undefined}
}%
\providecommand \@ifnum [1]{%
 \ifnum #1\expandafter \@firstoftwo
 \else \expandafter \@secondoftwo
 \fi
}%
\providecommand \@ifx [1]{%
 \ifx #1\expandafter \@firstoftwo
 \else \expandafter \@secondoftwo
 \fi
}%
\providecommand \natexlab [1]{#1}%
\providecommand \enquote  [1]{``#1''}%
\providecommand \bibnamefont  [1]{#1}%
\providecommand \bibfnamefont [1]{#1}%
\providecommand \citenamefont [1]{#1}%
\providecommand \href@noop [0]{\@secondoftwo}%
\providecommand \href [0]{\begingroup \@sanitize@url \@href}%
\providecommand \@href[1]{\@@startlink{#1}\@@href}%
\providecommand \@@href[1]{\endgroup#1\@@endlink}%
\providecommand \@sanitize@url [0]{\catcode `\\12\catcode `\$12\catcode
  `\&12\catcode `\#12\catcode `\^12\catcode `\_12\catcode `\%12\relax}%
\providecommand \@@startlink[1]{}%
\providecommand \@@endlink[0]{}%
\providecommand \url  [0]{\begingroup\@sanitize@url \@url }%
\providecommand \@url [1]{\endgroup\@href {#1}{\urlprefix }}%
\providecommand \urlprefix  [0]{URL }%
\providecommand \Eprint [0]{\href }%
\providecommand \doibase [0]{http://dx.doi.org/}%
\providecommand \selectlanguage [0]{\@gobble}%
\providecommand \bibinfo  [0]{\@secondoftwo}%
\providecommand \bibfield  [0]{\@secondoftwo}%
\providecommand \translation [1]{[#1]}%
\providecommand \BibitemOpen [0]{}%
\providecommand \bibitemStop [0]{}%
\providecommand \bibitemNoStop [0]{.\EOS\space}%
\providecommand \EOS [0]{\spacefactor3000\relax}%
\providecommand \BibitemShut  [1]{\csname bibitem#1\endcsname}%
\let\auto@bib@innerbib\@empty
%</preamble>
\bibitem [{\citenamefont {McCluskey}\ \emph
  {et~al.}(2019{\natexlab{a}})\citenamefont {McCluskey}, \citenamefont
  {Sanchez-Fernandez}, \citenamefont {Edler}, \citenamefont {Parker},
  \citenamefont {Jackson}, \citenamefont {Campbell},\ and\ \citenamefont
  {Arnold}}]{mccluskey_bayesian_2019}%
  \BibitemOpen
  \bibfield  {author} {\bibinfo {author} {\bibfnamefont {A.~R.}\ \bibnamefont
  {McCluskey}}, \bibinfo {author} {\bibfnamefont {A.}~\bibnamefont
  {Sanchez-Fernandez}}, \bibinfo {author} {\bibfnamefont {K.~J.}\ \bibnamefont
  {Edler}}, \bibinfo {author} {\bibfnamefont {S.~C.}\ \bibnamefont {Parker}},
  \bibinfo {author} {\bibfnamefont {A.~J.}\ \bibnamefont {Jackson}}, \bibinfo
  {author} {\bibfnamefont {R.~A.}\ \bibnamefont {Campbell}}, \ and\ \bibinfo
  {author} {\bibfnamefont {T.}~\bibnamefont {Arnold}},\ }\href {\doibase
  10.1039/C9CP00203K} {\bibfield  {journal} {\bibinfo  {journal} {Phys. Chem,
  Chem. Phys.}\ }\textbf {\bibinfo {volume} {21}},\ \bibinfo {pages} {6133}
  (\bibinfo {year} {2019}{\natexlab{a}})}\BibitemShut {NoStop}%
\bibitem [{\citenamefont {Clifton}\ \emph {et~al.}(2019)\citenamefont
  {Clifton}, \citenamefont {Paracini}, \citenamefont {Hughes}, \citenamefont
  {Lakey}, \citenamefont {Steinke}, \citenamefont {Cooper}, \citenamefont
  {Gavutis},\ and\ \citenamefont {Skoda}}]{clifton_self_2019}%
  \BibitemOpen
  \bibfield  {author} {\bibinfo {author} {\bibfnamefont {L.~A.}\ \bibnamefont
  {Clifton}}, \bibinfo {author} {\bibfnamefont {N.}~\bibnamefont {Paracini}},
  \bibinfo {author} {\bibfnamefont {A.~V.}\ \bibnamefont {Hughes}}, \bibinfo
  {author} {\bibfnamefont {J.~H.}\ \bibnamefont {Lakey}}, \bibinfo {author}
  {\bibfnamefont {N.-J.}\ \bibnamefont {Steinke}}, \bibinfo {author}
  {\bibfnamefont {J.~F.~K.}\ \bibnamefont {Cooper}}, \bibinfo {author}
  {\bibfnamefont {M.}~\bibnamefont {Gavutis}}, \ and\ \bibinfo {author}
  {\bibfnamefont {M.~W.~A.}\ \bibnamefont {Skoda}},\ }\href {\doibase
  10.1021/acs.langmuir.9b02350} {\bibfield  {journal} {\bibinfo  {journal}
  {Langmuir}\ }\textbf {\bibinfo {volume} {35}},\ \bibinfo {pages} {13735}
  (\bibinfo {year} {2019})}\BibitemShut {NoStop}%
\bibitem [{\citenamefont {P\'{e}rez}\ \emph {et~al.}(2019)\citenamefont
  {P\'{e}rez}, \citenamefont {Bernardo}, \citenamefont {Gaspar}, \citenamefont
  {Cooper}, \citenamefont {Bastianini}, \citenamefont {Parnell},\ and\
  \citenamefont {Dunbar}}]{perez_determination_2019}%
  \BibitemOpen
  \bibfield  {author} {\bibinfo {author} {\bibfnamefont {G.~E.}\ \bibnamefont
  {P\'{e}rez}}, \bibinfo {author} {\bibfnamefont {G.}~\bibnamefont {Bernardo}},
  \bibinfo {author} {\bibfnamefont {H.}~\bibnamefont {Gaspar}}, \bibinfo
  {author} {\bibfnamefont {J.~F.~K.}\ \bibnamefont {Cooper}}, \bibinfo {author}
  {\bibfnamefont {F.}~\bibnamefont {Bastianini}}, \bibinfo {author}
  {\bibfnamefont {A.~J.}\ \bibnamefont {Parnell}}, \ and\ \bibinfo {author}
  {\bibfnamefont {A.~D.~F.}\ \bibnamefont {Dunbar}},\ }\href {\doibase
  10.1021/acsami.9b02700} {\bibfield  {journal} {\bibinfo  {journal} {ACS Appl.
  Mater. Interfaces}\ }\textbf {\bibinfo {volume} {11}},\ \bibinfo {pages}
  {13803} (\bibinfo {year} {2019})}\BibitemShut {NoStop}%
\bibitem [{\citenamefont {Majkrzak}\ and\ \citenamefont
  {Berk}(1995{\natexlab{a}})}]{majkrzak_exact_1995}%
  \BibitemOpen
  \bibfield  {author} {\bibinfo {author} {\bibfnamefont {C.~F.}\ \bibnamefont
  {Majkrzak}}\ and\ \bibinfo {author} {\bibfnamefont {N.~F.}\ \bibnamefont
  {Berk}},\ }\href {\doibase 10.1103/PhysRevB.52.10827} {\bibfield  {journal}
  {\bibinfo  {journal} {Phys. Rev. B}\ }\textbf {\bibinfo {volume} {52}},\
  \bibinfo {pages} {10827} (\bibinfo {year} {1995}{\natexlab{a}})}\BibitemShut
  {NoStop}%
\bibitem [{\citenamefont {Kirby}\ \emph {et~al.}(2012)\citenamefont {Kirby},
  \citenamefont {Kienzle}, \citenamefont {Maranville}, \citenamefont {Berk},
  \citenamefont {Krycka}, \citenamefont {Heinrich},\ and\ \citenamefont
  {Majkrzak}}]{kirby_phase_2012}%
  \BibitemOpen
  \bibfield  {author} {\bibinfo {author} {\bibfnamefont {B.~J.}\ \bibnamefont
  {Kirby}}, \bibinfo {author} {\bibfnamefont {P.~A.}\ \bibnamefont {Kienzle}},
  \bibinfo {author} {\bibfnamefont {B.~B.}\ \bibnamefont {Maranville}},
  \bibinfo {author} {\bibfnamefont {N.~F.}\ \bibnamefont {Berk}}, \bibinfo
  {author} {\bibfnamefont {J.}~\bibnamefont {Krycka}}, \bibinfo {author}
  {\bibfnamefont {F.}~\bibnamefont {Heinrich}}, \ and\ \bibinfo {author}
  {\bibfnamefont {C.~F.}\ \bibnamefont {Majkrzak}},\ }\href {\doibase
  10.1016/j.cocis.2011.11.001} {\bibfield  {journal} {\bibinfo  {journal}
  {Curr. Opin. Colloid Interface Sci}\ }\textbf {\bibinfo {volume} {17}},\
  \bibinfo {pages} {44} (\bibinfo {year} {2012})}\BibitemShut {NoStop}%
\bibitem [{\citenamefont {de~Haan}\ \emph {et~al.}(1995)\citenamefont
  {de~Haan}, \citenamefont {van Well}, \citenamefont {Adenwalla},\ and\
  \citenamefont {Felcher}}]{haan_retrieval_1995}%
  \BibitemOpen
  \bibfield  {author} {\bibinfo {author} {\bibfnamefont {V.-O.}\ \bibnamefont
  {de~Haan}}, \bibinfo {author} {\bibfnamefont {A.~A.}\ \bibnamefont {van
  Well}}, \bibinfo {author} {\bibfnamefont {S.}~\bibnamefont {Adenwalla}}, \
  and\ \bibinfo {author} {\bibfnamefont {G.~P.}\ \bibnamefont {Felcher}},\
  }\href {\doibase 10.1103/PhysRevB.52.10831} {\bibfield  {journal} {\bibinfo
  {journal} {Phys. Rev. B}\ }\textbf {\bibinfo {volume} {52}},\ \bibinfo
  {pages} {10831} (\bibinfo {year} {1995})}\BibitemShut {NoStop}%
\bibitem [{\citenamefont {Majkrzak}\ \emph {et~al.}(2003)\citenamefont
  {Majkrzak}, \citenamefont {Berk},\ and\ \citenamefont
  {Perez-Salas}}]{majkrzak_phase_2003}%
  \BibitemOpen
  \bibfield  {author} {\bibinfo {author} {\bibfnamefont {C.~F.}\ \bibnamefont
  {Majkrzak}}, \bibinfo {author} {\bibfnamefont {N.~F.}\ \bibnamefont {Berk}},
  \ and\ \bibinfo {author} {\bibfnamefont {U.~A.}\ \bibnamefont
  {Perez-Salas}},\ }\href {\doibase 10.1021/la0341254} {\bibfield  {journal}
  {\bibinfo  {journal} {Phys. Rev. B}\ }\textbf {\bibinfo {volume} {19}},\
  \bibinfo {pages} {7796} (\bibinfo {year} {2003})}\BibitemShut {NoStop}%
\bibitem [{\citenamefont {Nikova}\ \emph {et~al.}(2019)\citenamefont {Nikova},
  \citenamefont {Salamatov}, \citenamefont {Kravtsov},\ and\ \citenamefont
  {Ustinov}}]{nikova_novel_2019}%
  \BibitemOpen
  \bibfield  {author} {\bibinfo {author} {\bibfnamefont {E.~S.}\ \bibnamefont
  {Nikova}}, \bibinfo {author} {\bibfnamefont {Y.~A.}\ \bibnamefont
  {Salamatov}}, \bibinfo {author} {\bibfnamefont {E.~A.}\ \bibnamefont
  {Kravtsov}}, \ and\ \bibinfo {author} {\bibfnamefont {V.~V.}\ \bibnamefont
  {Ustinov}},\ }\href {\doibase 10.1088/1742-6596/1389/1/012153} {\bibfield
  {journal} {\bibinfo  {journal} {J. Phys.: Conf. Ser.}\ }\textbf {\bibinfo
  {volume} {1389}},\ \bibinfo {pages} {012153} (\bibinfo {year}
  {2019})}\BibitemShut {NoStop}%
\bibitem [{\citenamefont {Leeb}\ \emph {et~al.}(1998)\citenamefont {Leeb},
  \citenamefont {Kasper},\ and\ \citenamefont
  {Lipperheide}}]{leeb_determination_1998}%
  \BibitemOpen
  \bibfield  {author} {\bibinfo {author} {\bibfnamefont {H.}~\bibnamefont
  {Leeb}}, \bibinfo {author} {\bibfnamefont {J.}~\bibnamefont {Kasper}}, \ and\
  \bibinfo {author} {\bibfnamefont {R.}~\bibnamefont {Lipperheide}},\ }\href
  {\doibase 10.1016/S0375-9601(97)00972-9} {\bibfield  {journal} {\bibinfo
  {journal} {Phys. Lett. A}\ }\textbf {\bibinfo {volume} {239}},\ \bibinfo
  {pages} {147} (\bibinfo {year} {1998})}\BibitemShut {NoStop}%
\bibitem [{\citenamefont {Fiedeldey}\ \emph {et~al.}(1992)\citenamefont
  {Fiedeldey}, \citenamefont {Lipperheide}, \citenamefont {Leeb},\ and\
  \citenamefont {Sofianos}}]{fiedeldey_proposal_1992}%
  \BibitemOpen
  \bibfield  {author} {\bibinfo {author} {\bibfnamefont {H.}~\bibnamefont
  {Fiedeldey}}, \bibinfo {author} {\bibfnamefont {R.}~\bibnamefont
  {Lipperheide}}, \bibinfo {author} {\bibfnamefont {H.}~\bibnamefont {Leeb}}, \
  and\ \bibinfo {author} {\bibfnamefont {S.~A.}\ \bibnamefont {Sofianos}},\
  }\href {\doibase 10.1016/0375-9601(92)90885-P} {\bibfield  {journal}
  {\bibinfo  {journal} {Phys. Lett. A}\ }\textbf {\bibinfo {volume} {170}},\
  \bibinfo {pages} {347} (\bibinfo {year} {1992})}\BibitemShut {NoStop}%
\bibitem [{\citenamefont {Majkrzak}\ and\ \citenamefont
  {Berk}(1995{\natexlab{b}})}]{majkrzak_exact_1998}%
  \BibitemOpen
  \bibfield  {author} {\bibinfo {author} {\bibfnamefont {C.~F.}\ \bibnamefont
  {Majkrzak}}\ and\ \bibinfo {author} {\bibfnamefont {N.~F.}\ \bibnamefont
  {Berk}},\ }\href {\doibase 10.1103/PhysRevB.58.15416} {\bibfield  {journal}
  {\bibinfo  {journal} {Phys. Rev. B}\ }\textbf {\bibinfo {volume} {58}},\
  \bibinfo {pages} {15416} (\bibinfo {year} {1995}{\natexlab{b}})}\BibitemShut
  {NoStop}%
\bibitem [{\citenamefont {Majkrzak}\ \emph {et~al.}(2000)\citenamefont
  {Majkrzak}, \citenamefont {Berk}, \citenamefont {Kreuger}, \citenamefont
  {Dura}, \citenamefont {Tarek}, \citenamefont {Tobias}, \citenamefont {Silin},
  \citenamefont {Meuse},\ and\ \citenamefont {Plant}}]{majkrzak_first_2000}%
  \BibitemOpen
  \bibfield  {author} {\bibinfo {author} {\bibfnamefont {C.~F.}\ \bibnamefont
  {Majkrzak}}, \bibinfo {author} {\bibfnamefont {N.~F.}\ \bibnamefont {Berk}},
  \bibinfo {author} {\bibfnamefont {S.}~\bibnamefont {Kreuger}}, \bibinfo
  {author} {\bibfnamefont {J.~A.}\ \bibnamefont {Dura}}, \bibinfo {author}
  {\bibfnamefont {M.}~\bibnamefont {Tarek}}, \bibinfo {author} {\bibfnamefont
  {D.}~\bibnamefont {Tobias}}, \bibinfo {author} {\bibfnamefont
  {V.}~\bibnamefont {Silin}}, \bibinfo {author} {\bibfnamefont
  {J.}~\bibnamefont {Meuse}, \bibfnamefont {C.~W.~Woodward}}, \ and\ \bibinfo
  {author} {\bibfnamefont {A.~L.}\ \bibnamefont {Plant}},\ }\href {\doibase
  10.1016/S0006-3495(00)76564-7} {\bibfield  {journal} {\bibinfo  {journal}
  {Biophys. J.}\ }\textbf {\bibinfo {volume} {79}},\ \bibinfo {pages} {3330}
  (\bibinfo {year} {2000})}\BibitemShut {NoStop}%
\bibitem [{\citenamefont {Koutsioubas}(2019)}]{koutsioubas_model_2019}%
  \BibitemOpen
  \bibfield  {author} {\bibinfo {author} {\bibfnamefont {A.}~\bibnamefont
  {Koutsioubas}},\ }\href {\doibase 10.1107/S1600576719003534} {\bibfield
  {journal} {\bibinfo  {journal} {J. Appl. Crystallogr.}\ }\textbf {\bibinfo
  {volume} {52}},\ \bibinfo {pages} {538} (\bibinfo {year} {2019})}\BibitemShut
  {NoStop}%
\bibitem [{\citenamefont {Pedersen}\ and\ \citenamefont
  {Hamley}(1994)}]{pedersen_analysis_1994}%
  \BibitemOpen
  \bibfield  {author} {\bibinfo {author} {\bibfnamefont {J.~S.}\ \bibnamefont
  {Pedersen}}\ and\ \bibinfo {author} {\bibfnamefont {I.~W.}\ \bibnamefont
  {Hamley}},\ }\href {\doibase 10.1016/0921-4526(94)90117-1} {\bibfield
  {journal} {\bibinfo  {journal} {Physica B}\ }\textbf {\bibinfo {volume}
  {198}},\ \bibinfo {pages} {16} (\bibinfo {year} {1994})}\BibitemShut
  {NoStop}%
\bibitem [{\citenamefont {de~Haan}\ and\ \citenamefont
  {Drijkoningen}(1994)}]{haan_genetic_1994}%
  \BibitemOpen
  \bibfield  {author} {\bibinfo {author} {\bibfnamefont {V.-O.}\ \bibnamefont
  {de~Haan}}\ and\ \bibinfo {author} {\bibfnamefont {G.~G.}\ \bibnamefont
  {Drijkoningen}},\ }\href {\doibase 10.1016/0921-4526(94)90118-X} {\bibfield
  {journal} {\bibinfo  {journal} {Physica B}\ }\textbf {\bibinfo {volume}
  {198}},\ \bibinfo {pages} {24} (\bibinfo {year} {1994})}\BibitemShut
  {NoStop}%
\bibitem [{\citenamefont {Nelson}(2006)}]{nelson_motofit_2006}%
  \BibitemOpen
  \bibfield  {author} {\bibinfo {author} {\bibfnamefont {A.~R.~J.}\
  \bibnamefont {Nelson}},\ }\href {\doibase 10.1107/S0021889806005073}
  {\bibfield  {journal} {\bibinfo  {journal} {J. Appl. Crystallogr.}\ }\textbf
  {\bibinfo {volume} {39}},\ \bibinfo {pages} {273} (\bibinfo {year}
  {2006})}\BibitemShut {NoStop}%
\bibitem [{\citenamefont {van~der Lee}\ \emph {et~al.}(2007)\citenamefont
  {van~der Lee}, \citenamefont {Salah},\ and\ \citenamefont
  {Harzallah}}]{lee_comparison_2007}%
  \BibitemOpen
  \bibfield  {author} {\bibinfo {author} {\bibfnamefont {A.}~\bibnamefont
  {van~der Lee}}, \bibinfo {author} {\bibfnamefont {F.}~\bibnamefont {Salah}},
  \ and\ \bibinfo {author} {\bibfnamefont {B.}~\bibnamefont {Harzallah}},\
  }\href {\doibase 10.1107/S0021889807032207} {\bibfield  {journal} {\bibinfo
  {journal} {J. Appl. Crystallogr.}\ }\textbf {\bibinfo {volume} {40}},\
  \bibinfo {pages} {820} (\bibinfo {year} {2007})}\BibitemShut {NoStop}%
\bibitem [{\citenamefont {Gerelli}(2016{\natexlab{a}})}]{gerelli_aurore_2016}%
  \BibitemOpen
  \bibfield  {author} {\bibinfo {author} {\bibfnamefont {Y.}~\bibnamefont
  {Gerelli}},\ }\href {\doibase 10.1107/S1600576716000108} {\bibfield
  {journal} {\bibinfo  {journal} {J. Appl. Crystallogr.}\ }\textbf {\bibinfo
  {volume} {49}},\ \bibinfo {pages} {330} (\bibinfo {year}
  {2016}{\natexlab{a}})}\BibitemShut {NoStop}%
\bibitem [{\citenamefont {Gerelli}(2016{\natexlab{b}})}]{gerelli_aurore_2016b}%
  \BibitemOpen
  \bibfield  {author} {\bibinfo {author} {\bibfnamefont {Y.}~\bibnamefont
  {Gerelli}},\ }\href {\doibase 10.1107/S1600576716002466} {\bibfield
  {journal} {\bibinfo  {journal} {J. Appl. Crystallogr.}\ }\textbf {\bibinfo
  {volume} {49}},\ \bibinfo {pages} {712} (\bibinfo {year}
  {2016}{\natexlab{b}})}\BibitemShut {NoStop}%
\bibitem [{\citenamefont {Nelson}\ and\ \citenamefont
  {Prescott}(2019)}]{nelson_refnx_2019}%
  \BibitemOpen
  \bibfield  {author} {\bibinfo {author} {\bibfnamefont {A.~R.~J.}\
  \bibnamefont {Nelson}}\ and\ \bibinfo {author} {\bibfnamefont {S.~W.}\
  \bibnamefont {Prescott}},\ }\href {\doibase 10.1107/S1600576718017296}
  {\bibfield  {journal} {\bibinfo  {journal} {J. Appl. Crystallogr.}\ }\textbf
  {\bibinfo {volume} {52}},\ \bibinfo {pages} {193} (\bibinfo {year}
  {2019})}\BibitemShut {NoStop}%
\bibitem [{\citenamefont {Abel\`{e}s}(1948)}]{abeles_propagation_1948}%
  \BibitemOpen
  \bibfield  {author} {\bibinfo {author} {\bibfnamefont {F.}~\bibnamefont
  {Abel\`{e}s}},\ }\href {\doibase 10.1051/anphys/194812030504} {\bibfield
  {journal} {\bibinfo  {journal} {Ann. Phys.}\ }\textbf {\bibinfo {volume}
  {12}},\ \bibinfo {pages} {504} (\bibinfo {year} {1948})}\BibitemShut
  {NoStop}%
\bibitem [{\citenamefont {Parratt}(1954)}]{parratt_surface_1954}%
  \BibitemOpen
  \bibfield  {author} {\bibinfo {author} {\bibfnamefont {L.~G.}\ \bibnamefont
  {Parratt}},\ }\href {\doibase 10.1103/PhysRev.95.359} {\bibfield  {journal}
  {\bibinfo  {journal} {Phys. Rev.}\ }\textbf {\bibinfo {volume} {95}},\
  \bibinfo {pages} {359} (\bibinfo {year} {1954})}\BibitemShut {NoStop}%
\bibitem [{\citenamefont {Klibanov}\ and\ \citenamefont
  {Sacks}(1992)}]{klibanov_phaseless_1992}%
  \BibitemOpen
  \bibfield  {author} {\bibinfo {author} {\bibfnamefont {M.~V.}\ \bibnamefont
  {Klibanov}}\ and\ \bibinfo {author} {\bibfnamefont {P.~E.}\ \bibnamefont
  {Sacks}},\ }\href {\doibase 10.1063/1.529990} {\bibfield  {journal} {\bibinfo
   {journal} {J. Math. Phys.}\ }\textbf {\bibinfo {volume} {33}},\ \bibinfo
  {pages} {3813} (\bibinfo {year} {1992})}\BibitemShut {NoStop}%
\bibitem [{\citenamefont {Klibanov}\ and\ \citenamefont
  {Sacks}(1994)}]{klibanov_use_1994}%
  \BibitemOpen
  \bibfield  {author} {\bibinfo {author} {\bibfnamefont {M.~V.}\ \bibnamefont
  {Klibanov}}\ and\ \bibinfo {author} {\bibfnamefont {P.~E.}\ \bibnamefont
  {Sacks}},\ }\href {\doibase 10.1006/jcph.1994.1099} {\bibfield  {journal}
  {\bibinfo  {journal} {J. Comp. Phys.}\ }\textbf {\bibinfo {volume} {112}},\
  \bibinfo {pages} {273} (\bibinfo {year} {1994})}\BibitemShut {NoStop}%
\bibitem [{\citenamefont {Waldie}\ \emph {et~al.}(2018)\citenamefont {Waldie},
  \citenamefont {Lind}, \citenamefont {Browning}, \citenamefont {Moulin},
  \citenamefont {Haertlein}, \citenamefont {Forsyth}, \citenamefont {Luchini},
  \citenamefont {Strohmeier}, \citenamefont {Pichler}, \citenamefont {Maric},\
  and\ \citenamefont {C\'{a}rdenas}}]{waldie_localization_2018}%
  \BibitemOpen
  \bibfield  {author} {\bibinfo {author} {\bibfnamefont {S.}~\bibnamefont
  {Waldie}}, \bibinfo {author} {\bibfnamefont {T.~K.}\ \bibnamefont {Lind}},
  \bibinfo {author} {\bibfnamefont {K.}~\bibnamefont {Browning}}, \bibinfo
  {author} {\bibfnamefont {M.}~\bibnamefont {Moulin}}, \bibinfo {author}
  {\bibfnamefont {M.}~\bibnamefont {Haertlein}}, \bibinfo {author}
  {\bibfnamefont {V.~T.}\ \bibnamefont {Forsyth}}, \bibinfo {author}
  {\bibfnamefont {A.}~\bibnamefont {Luchini}}, \bibinfo {author} {\bibfnamefont
  {G.~A.}\ \bibnamefont {Strohmeier}}, \bibinfo {author} {\bibfnamefont
  {H.}~\bibnamefont {Pichler}}, \bibinfo {author} {\bibfnamefont
  {S.}~\bibnamefont {Maric}}, \ and\ \bibinfo {author} {\bibfnamefont
  {M.}~\bibnamefont {C\'{a}rdenas}},\ }\href {\doibase
  10.1021/acs.langmuir.7b02716} {\bibfield  {journal} {\bibinfo  {journal}
  {Langmuir}\ }\textbf {\bibinfo {volume} {34}},\ \bibinfo {pages} {472}
  (\bibinfo {year} {2018})}\BibitemShut {NoStop}%
\bibitem [{\citenamefont {Campbell}\ \emph {et~al.}(2018)\citenamefont
  {Campbell}, \citenamefont {Saaka}, \citenamefont {Shao}, \citenamefont
  {Gerelli}, \citenamefont {Cubitt}, \citenamefont {Nazaruk}, \citenamefont
  {Matyszewska},\ and\ \citenamefont {Lawrence}}]{campbell_structure_2018}%
  \BibitemOpen
  \bibfield  {author} {\bibinfo {author} {\bibfnamefont {R.~A.}\ \bibnamefont
  {Campbell}}, \bibinfo {author} {\bibfnamefont {Y.}~\bibnamefont {Saaka}},
  \bibinfo {author} {\bibfnamefont {Y.}~\bibnamefont {Shao}}, \bibinfo {author}
  {\bibfnamefont {Y.}~\bibnamefont {Gerelli}}, \bibinfo {author} {\bibfnamefont
  {R.}~\bibnamefont {Cubitt}}, \bibinfo {author} {\bibfnamefont
  {E.}~\bibnamefont {Nazaruk}}, \bibinfo {author} {\bibfnamefont
  {D.}~\bibnamefont {Matyszewska}}, \ and\ \bibinfo {author} {\bibfnamefont
  {M.~J.}\ \bibnamefont {Lawrence}},\ }\href {\doibase
  10.1016/j.jcis.2018.07.022} {\bibfield  {journal} {\bibinfo  {journal} {J.
  Colloid Interface Sci.}\ }\textbf {\bibinfo {volume} {531}},\ \bibinfo
  {pages} {98} (\bibinfo {year} {2018})}\BibitemShut {NoStop}%
\bibitem [{\citenamefont {Schalke}\ and\ \citenamefont
  {L\"{o}sche}(2000)}]{schalke_structural_2000}%
  \BibitemOpen
  \bibfield  {author} {\bibinfo {author} {\bibfnamefont {M.}~\bibnamefont
  {Schalke}}\ and\ \bibinfo {author} {\bibfnamefont {M.}~\bibnamefont
  {L\"{o}sche}},\ }\href {\doibase 10.1016/S0001-8686(00)00047-6} {\bibfield
  {journal} {\bibinfo  {journal} {Adv. Colloid Interface Sci.}\ }\textbf
  {\bibinfo {volume} {88}},\ \bibinfo {pages} {243} (\bibinfo {year}
  {2000})}\BibitemShut {NoStop}%
\bibitem [{\citenamefont {Heinrich}\ and\ \citenamefont
  {L\"{o}sche}(2014)}]{heinrich_zooming_2014}%
  \BibitemOpen
  \bibfield  {author} {\bibinfo {author} {\bibfnamefont {F.}~\bibnamefont
  {Heinrich}}\ and\ \bibinfo {author} {\bibfnamefont {M.}~\bibnamefont
  {L\"{o}sche}},\ }\href {\doibase 10.1016/j.bbamem.2014.03.007} {\bibfield
  {journal} {\bibinfo  {journal} {BBA-Biomembranes}\ }\textbf {\bibinfo
  {volume} {1838}},\ \bibinfo {pages} {2341} (\bibinfo {year}
  {2014})}\BibitemShut {NoStop}%
\bibitem [{\citenamefont {Heinrich}(2016)}]{heinrich_deuteration_2016}%
  \BibitemOpen
  \bibfield  {author} {\bibinfo {author} {\bibfnamefont {F.}~\bibnamefont
  {Heinrich}},\ }in\ \href {\doibase 10.1016/bs.mie.2015.05.019} {\emph
  {\bibinfo {booktitle} {Isotope Labeling of Biomolecules - Applications}}},\
  \bibinfo {series} {Methods in Enzymology}, Vol.\ \bibinfo {volume} {566},\
  \bibinfo {editor} {edited by\ \bibinfo {editor} {\bibfnamefont
  {Z.}~\bibnamefont {Kelman}}}\ (\bibinfo  {publisher} {Academic Press},\
  \bibinfo {year} {2016})\ pp.\ \bibinfo {pages} {211--230}\BibitemShut
  {NoStop}%
\bibitem [{\citenamefont {Sun}\ \emph {et~al.}(1994)\citenamefont {Sun},
  \citenamefont {Suter}, \citenamefont {Knewtson}, \citenamefont {Worthington},
  \citenamefont {Tristram-Nagle}, \citenamefont {Zhang},\ and\ \citenamefont
  {Nagle}}]{sun_order_1994}%
  \BibitemOpen
  \bibfield  {author} {\bibinfo {author} {\bibfnamefont {W.-J.}\ \bibnamefont
  {Sun}}, \bibinfo {author} {\bibfnamefont {R.~M.}\ \bibnamefont {Suter}},
  \bibinfo {author} {\bibfnamefont {M.~A.}\ \bibnamefont {Knewtson}}, \bibinfo
  {author} {\bibfnamefont {C.~R.}\ \bibnamefont {Worthington}}, \bibinfo
  {author} {\bibfnamefont {S.}~\bibnamefont {Tristram-Nagle}}, \bibinfo
  {author} {\bibfnamefont {R.}~\bibnamefont {Zhang}}, \ and\ \bibinfo {author}
  {\bibfnamefont {J.~F.}\ \bibnamefont {Nagle}},\ }\href {\doibase
  10.1103/PhysRevE.49.4665} {\bibfield  {journal} {\bibinfo  {journal} {Phys.
  Rev. Lett.}\ }\textbf {\bibinfo {volume} {49}},\ \bibinfo {pages} {4665}
  (\bibinfo {year} {1994})}\BibitemShut {NoStop}%
\bibitem [{\citenamefont {Armen}\ \emph {et~al.}(1998)\citenamefont {Armen},
  \citenamefont {Uitto},\ and\ \citenamefont
  {Feller}}]{armen_phospholipid_1998}%
  \BibitemOpen
  \bibfield  {author} {\bibinfo {author} {\bibfnamefont {R.~S.}\ \bibnamefont
  {Armen}}, \bibinfo {author} {\bibfnamefont {O.~D.}\ \bibnamefont {Uitto}}, \
  and\ \bibinfo {author} {\bibfnamefont {S.~E.}\ \bibnamefont {Feller}},\
  }\href {\doibase 10.1016/S0006-3495(98)77563-0} {\bibfield  {journal}
  {\bibinfo  {journal} {Biophys. J.}\ }\textbf {\bibinfo {volume} {75}},\
  \bibinfo {pages} {734} (\bibinfo {year} {1998})}\BibitemShut {NoStop}%
\bibitem [{\citenamefont {Hughes}(2017)}]{hughes_rascal_2017}%
  \BibitemOpen
  \bibfield  {author} {\bibinfo {author} {\bibfnamefont {A.~V.}\ \bibnamefont
  {Hughes}},\ }\href@noop {} {\enquote {\bibinfo {title} {Rascal
  sourceforge},}\ }\bibinfo {howpublished}
  {\url{https://sourceforge.net/projects/rscl/}} (\bibinfo {year} {2017}),\
  \bibinfo {note} {accessed: 2019-10-09}\BibitemShut {NoStop}%
\bibitem [{\citenamefont {Kienzle}\ \emph {et~al.}(2017)\citenamefont
  {Kienzle}, \citenamefont {Maranville}, \citenamefont {O'Donovan},
  \citenamefont {Ankner}, \citenamefont {Berk},\ and\ \citenamefont
  {Majkrzak}}]{kienzle_ncnr_2017}%
  \BibitemOpen
  \bibfield  {author} {\bibinfo {author} {\bibfnamefont {P.~A.}\ \bibnamefont
  {Kienzle}}, \bibinfo {author} {\bibfnamefont {B.~B.}\ \bibnamefont
  {Maranville}}, \bibinfo {author} {\bibfnamefont {K.~V.}\ \bibnamefont
  {O'Donovan}}, \bibinfo {author} {\bibfnamefont {J.~F.}\ \bibnamefont
  {Ankner}}, \bibinfo {author} {\bibfnamefont {N.~K.}\ \bibnamefont {Berk}}, \
  and\ \bibinfo {author} {\bibfnamefont {C.~F.}\ \bibnamefont {Majkrzak}},\
  }\href@noop {} {\enquote {\bibinfo {title} {{NCNR} reflectometry software},}\
  }\bibinfo {howpublished}
  {\url{https://www.nist.gov/ncnr/reflectometry-software}} (\bibinfo {year}
  {2017}),\ \bibinfo {note} {accessed: 2019-09-15}\BibitemShut {NoStop}%
\bibitem [{\citenamefont {Bayes}(1763)}]{bayes_essay_1763}%
  \BibitemOpen
  \bibfield  {author} {\bibinfo {author} {\bibfnamefont {T.}~\bibnamefont
  {Bayes}},\ }\href {\doibase 10.1098/rstl.1763.0053} {\bibfield  {journal}
  {\bibinfo  {journal} {Phil. Trans. Roy. Soc.}\ }\textbf {\bibinfo {volume}
  {53}},\ \bibinfo {pages} {370} (\bibinfo {year} {1763})}\BibitemShut
  {NoStop}%
\bibitem [{\citenamefont {Lee}\ and\ \citenamefont
  {Ditko}(1962)}]{lee_introducting_1962}%
  \BibitemOpen
  \bibfield  {author} {\bibinfo {author} {\bibfnamefont {S.}~\bibnamefont
  {Lee}}\ and\ \bibinfo {author} {\bibfnamefont {S.}~\bibnamefont {Ditko}},\
  }\href@noop {} {\bibfield  {journal} {\bibinfo  {journal} {Amazing Fantasy
  (Marvel Comics)}\ }\textbf {\bibinfo {volume} {15}} (\bibinfo {year}
  {1962})}\BibitemShut {NoStop}%
\bibitem [{\citenamefont {Mayer}\ \emph {et~al.}(2010)\citenamefont {Mayer},
  \citenamefont {Khairy},\ and\ \citenamefont {Howard}}]{mayer_drawing_2010}%
  \BibitemOpen
  \bibfield  {author} {\bibinfo {author} {\bibfnamefont {J.}~\bibnamefont
  {Mayer}}, \bibinfo {author} {\bibfnamefont {K.}~\bibnamefont {Khairy}}, \
  and\ \bibinfo {author} {\bibfnamefont {J.}~\bibnamefont {Howard}},\ }\href
  {\doibase 10.1119/1.3254017} {\bibfield  {journal} {\bibinfo  {journal} {Am.
  J. Phys.}\ }\textbf {\bibinfo {volume} {78}},\ \bibinfo {pages} {648}
  (\bibinfo {year} {2010})}\BibitemShut {NoStop}%
\bibitem [{\citenamefont {Sivia}\ \emph {et~al.}(1991)\citenamefont {Sivia},
  \citenamefont {Hamilton},\ and\ \citenamefont {Smith}}]{sivia_analysis_1991}%
  \BibitemOpen
  \bibfield  {author} {\bibinfo {author} {\bibfnamefont {D.~S.}\ \bibnamefont
  {Sivia}}, \bibinfo {author} {\bibfnamefont {W.~A.}\ \bibnamefont {Hamilton}},
  \ and\ \bibinfo {author} {\bibfnamefont {G.~S.}\ \bibnamefont {Smith}},\
  }\href {\doibase 10.1016/0921-4526(91)90042-D} {\bibfield  {journal}
  {\bibinfo  {journal} {Physica B}\ }\textbf {\bibinfo {volume} {173}},\
  \bibinfo {pages} {121} (\bibinfo {year} {1991})}\BibitemShut {NoStop}%
\bibitem [{\citenamefont {Geoghagan}\ \emph {et~al.}(1996)\citenamefont
  {Geoghagan}, \citenamefont {Jones}, \citenamefont {Sivia}, \citenamefont
  {Penfold},\ and\ \citenamefont {Clough}}]{geoghegan_experimental_1996}%
  \BibitemOpen
  \bibfield  {author} {\bibinfo {author} {\bibfnamefont {M.}~\bibnamefont
  {Geoghagan}}, \bibinfo {author} {\bibfnamefont {R.~A.~L.}\ \bibnamefont
  {Jones}}, \bibinfo {author} {\bibfnamefont {D.~S.}\ \bibnamefont {Sivia}},
  \bibinfo {author} {\bibfnamefont {J.}~\bibnamefont {Penfold}}, \ and\
  \bibinfo {author} {\bibfnamefont {A.~S.}\ \bibnamefont {Clough}},\ }\href
  {\doibase 10.1103/PhysRevE.53.825} {\bibfield  {journal} {\bibinfo  {journal}
  {Thin Solid Films}\ }\textbf {\bibinfo {volume} {53}},\ \bibinfo {pages}
  {825} (\bibinfo {year} {1996})}\BibitemShut {NoStop}%
\bibitem [{\citenamefont {Sivia}\ and\ \citenamefont
  {Webster}(1998)}]{sivia_bayesian_1998}%
  \BibitemOpen
  \bibfield  {author} {\bibinfo {author} {\bibfnamefont {D.~S.}\ \bibnamefont
  {Sivia}}\ and\ \bibinfo {author} {\bibfnamefont {J.~R.~P.}\ \bibnamefont
  {Webster}},\ }\href {\doibase 10.1016/S0921-4526(98)00259-2} {\bibfield
  {journal} {\bibinfo  {journal} {Physica B}\ }\textbf {\bibinfo {volume}
  {248}},\ \bibinfo {pages} {327} (\bibinfo {year} {1998})}\BibitemShut
  {NoStop}%
\bibitem [{\citenamefont {Hughes}\ \emph {et~al.}(2019)\citenamefont {Hughes},
  \citenamefont {Patel}, \citenamefont {Widmalm}, \citenamefont {Klauda},
  \citenamefont {Clifton},\ and\ \citenamefont {Im}}]{hughes_physical_2019}%
  \BibitemOpen
  \bibfield  {author} {\bibinfo {author} {\bibfnamefont {A.~V.}\ \bibnamefont
  {Hughes}}, \bibinfo {author} {\bibfnamefont {D.~S.}\ \bibnamefont {Patel}},
  \bibinfo {author} {\bibfnamefont {G.}~\bibnamefont {Widmalm}}, \bibinfo
  {author} {\bibfnamefont {J.~B.}\ \bibnamefont {Klauda}}, \bibinfo {author}
  {\bibfnamefont {L.~A.}\ \bibnamefont {Clifton}}, \ and\ \bibinfo {author}
  {\bibfnamefont {W.}~\bibnamefont {Im}},\ }\href {\doibase
  10.1016/j.bpj.2019.02.001} {\bibfield  {journal} {\bibinfo  {journal}
  {Biophys. J.}\ }\textbf {\bibinfo {volume} {116}},\ \bibinfo {pages} {1095}
  (\bibinfo {year} {2019})}\BibitemShut {NoStop}%
\bibitem [{Note1()}]{Note1}%
  \BibitemOpen
  \bibinfo {note} {In this work, a reduced notation has been used, where the
  background information parameter, $I$ is implied.}\BibitemShut {Stop}%
\bibitem [{\citenamefont {Pullen}\ and\ \citenamefont
  {Morris}(2014)}]{pullen_bayesian_2014}%
  \BibitemOpen
  \bibfield  {author} {\bibinfo {author} {\bibfnamefont {N.}~\bibnamefont
  {Pullen}}\ and\ \bibinfo {author} {\bibfnamefont {R.~J.}\ \bibnamefont
  {Morris}},\ }\href {\doibase 10.1371/journal.pone.0088419} {\bibfield
  {journal} {\bibinfo  {journal} {PLOS ONE}\ }\textbf {\bibinfo {volume} {9}},\
  \bibinfo {pages} {e88419} (\bibinfo {year} {2014})}\BibitemShut {NoStop}%
\bibitem [{\citenamefont {Skilling}(2006)}]{skilling_nested_2006}%
  \BibitemOpen
  \bibfield  {author} {\bibinfo {author} {\bibfnamefont {J.}~\bibnamefont
  {Skilling}},\ }\href {\doibase 10.1063/1.1835238} {\bibfield  {journal}
  {\bibinfo  {journal} {AIP Conference Proceedings}\ }\textbf {\bibinfo
  {volume} {735}},\ \bibinfo {pages} {395} (\bibinfo {year}
  {2006})}\BibitemShut {NoStop}%
\bibitem [{\citenamefont {Speagle}(2019)}]{speagle_dynesty_2019}%
  \BibitemOpen
  \bibfield  {author} {\bibinfo {author} {\bibfnamefont {J.~S.}\ \bibnamefont
  {Speagle}},\ }\href@noop {} {\enquote {\bibinfo {title} {dynesty: {A}
  {D}ynamic {N}ested {S}ampling {P}ackage for {E}stimating {B}ayesian
  {P}osteriors and {E}vidences},}\ } (\bibinfo {year} {2019}),\ \Eprint
  {http://arxiv.org/abs/arXiv:1904.02180} {arXiv:1904.02180} \BibitemShut
  {NoStop}%
\bibitem [{\citenamefont {Sivia}\ and\ \citenamefont
  {Skilling}(2005)}]{sivia_data_2005}%
  \BibitemOpen
  \bibfield  {author} {\bibinfo {author} {\bibfnamefont {D.~S.}\ \bibnamefont
  {Sivia}}\ and\ \bibinfo {author} {\bibfnamefont {J.}~\bibnamefont
  {Skilling}},\ }\href@noop {} {\emph {\bibinfo {title} {Data Analysis: A
  Bayesian Tutorial}}},\ \bibinfo {edition} {2nd}\ ed.\ (\bibinfo  {publisher}
  {{Oxford University Press}},\ \bibinfo {address} {Oxford, UK},\ \bibinfo
  {year} {2005})\BibitemShut {NoStop}%
\bibitem [{\citenamefont {Nelson}\ \emph {et~al.}(2019)\citenamefont {Nelson},
  \citenamefont {Prescott}, \citenamefont {Hannum}, \citenamefont {McCluskey},\
  and\ \citenamefont {Gresham}}]{refnx_0.1.7_2019}%
  \BibitemOpen
  \bibfield  {author} {\bibinfo {author} {\bibfnamefont {A.}~\bibnamefont
  {Nelson}}, \bibinfo {author} {\bibfnamefont {S.}~\bibnamefont {Prescott}},
  \bibinfo {author} {\bibfnamefont {A.}~\bibnamefont {Hannum}}, \bibinfo
  {author} {\bibfnamefont {A.~R.}\ \bibnamefont {McCluskey}}, \ and\ \bibinfo
  {author} {\bibfnamefont {I.}~\bibnamefont {Gresham}},\ }\href {\doibase
  10.5281/zenodo.3240935} {\enquote {\bibinfo {title} {refnx/refnx v0.1.7},}\ }
  (\bibinfo {year} {2019})\BibitemShut {NoStop}%
\bibitem [{\citenamefont {Kass}\ and\ \citenamefont
  {Raftery}(1995)}]{kass_bayes_1995}%
  \BibitemOpen
  \bibfield  {author} {\bibinfo {author} {\bibfnamefont {R.~E.}\ \bibnamefont
  {Kass}}\ and\ \bibinfo {author} {\bibfnamefont {A.~E.}\ \bibnamefont
  {Raftery}},\ }\href {\doibase 10.2307/2291091} {\bibfield  {journal}
  {\bibinfo  {journal} {J. Am. Stat. Assoc.}\ }\textbf {\bibinfo {volume}
  {90}},\ \bibinfo {pages} {773} (\bibinfo {year} {1995})}\BibitemShut
  {NoStop}%
\bibitem [{\citenamefont {Cornish}\ and\ \citenamefont
  {Littenberg}(2007)}]{cornish_tests_2007}%
  \BibitemOpen
  \bibfield  {author} {\bibinfo {author} {\bibfnamefont {N.~J.}\ \bibnamefont
  {Cornish}}\ and\ \bibinfo {author} {\bibfnamefont {T.~B.}\ \bibnamefont
  {Littenberg}},\ }\href {\doibase 10.1103/PhysRevD.76.083006} {\bibfield
  {journal} {\bibinfo  {journal} {Phys. Rev. D}\ }\textbf {\bibinfo {volume}
  {76}},\ \bibinfo {pages} {083006} (\bibinfo {year} {2007})}\BibitemShut
  {NoStop}%
\bibitem [{\citenamefont {Ensign}\ and\ \citenamefont
  {Pande}(2010)}]{ensign_bayesian_2010}%
  \BibitemOpen
  \bibfield  {author} {\bibinfo {author} {\bibfnamefont {D.~L.}\ \bibnamefont
  {Ensign}}\ and\ \bibinfo {author} {\bibfnamefont {V.~S.}\ \bibnamefont
  {Pande}},\ }\href {\doibase 10.1021/jp906786b} {\bibfield  {journal}
  {\bibinfo  {journal} {J. Phys. Chem. B}\ }\textbf {\bibinfo {volume} {114}},\
  \bibinfo {pages} {280} (\bibinfo {year} {2010})}\BibitemShut {NoStop}%
\bibitem [{\citenamefont {Hollinshead}\ \emph {et~al.}(2009)\citenamefont
  {Hollinshead}, \citenamefont {Harvey}, \citenamefont {Barlow}, \citenamefont
  {Webster}, \citenamefont {Hughes}, \citenamefont {Weston},\ and\
  \citenamefont {Lawrence}}]{hollinshead_effects_2009}%
  \BibitemOpen
  \bibfield  {author} {\bibinfo {author} {\bibfnamefont {C.~M.}\ \bibnamefont
  {Hollinshead}}, \bibinfo {author} {\bibfnamefont {R.~D.}\ \bibnamefont
  {Harvey}}, \bibinfo {author} {\bibfnamefont {D.~J.}\ \bibnamefont {Barlow}},
  \bibinfo {author} {\bibfnamefont {J.~R.~P.}\ \bibnamefont {Webster}},
  \bibinfo {author} {\bibfnamefont {A.~V.}\ \bibnamefont {Hughes}}, \bibinfo
  {author} {\bibfnamefont {A.}~\bibnamefont {Weston}}, \ and\ \bibinfo {author}
  {\bibfnamefont {M.~J.}\ \bibnamefont {Lawrence}},\ }\href {\doibase
  10.1021/la8028319} {\bibfield  {journal} {\bibinfo  {journal} {Langmuir}\
  }\textbf {\bibinfo {volume} {25}},\ \bibinfo {pages} {4070} (\bibinfo {year}
  {2009})}\BibitemShut {NoStop}%
\bibitem [{\citenamefont {McCluskey}\ \emph
  {et~al.}(2019{\natexlab{b}})\citenamefont {McCluskey}, \citenamefont {Grant},
  \citenamefont {Smith}, \citenamefont {Rawle}, \citenamefont {Barlow},
  \citenamefont {Lawrence}, \citenamefont {Parker},\ and\ \citenamefont
  {Edler}}]{mccluskey_assessing_2019}%
  \BibitemOpen
  \bibfield  {author} {\bibinfo {author} {\bibfnamefont {A.~R.}\ \bibnamefont
  {McCluskey}}, \bibinfo {author} {\bibfnamefont {J.}~\bibnamefont {Grant}},
  \bibinfo {author} {\bibfnamefont {A.~J.}\ \bibnamefont {Smith}}, \bibinfo
  {author} {\bibfnamefont {J.~L.}\ \bibnamefont {Rawle}}, \bibinfo {author}
  {\bibfnamefont {D.~J.}\ \bibnamefont {Barlow}}, \bibinfo {author}
  {\bibfnamefont {M.~J.}\ \bibnamefont {Lawrence}}, \bibinfo {author}
  {\bibfnamefont {S.~C.}\ \bibnamefont {Parker}}, \ and\ \bibinfo {author}
  {\bibfnamefont {K.~J.}\ \bibnamefont {Edler}},\ }\href {\doibase
  10.1088/2399-6528/ab12a9} {\bibfield  {journal} {\bibinfo  {journal} {J.
  Phys. Comm.}\ }\textbf {\bibinfo {volume} {3}},\ \bibinfo {pages} {075001}
  (\bibinfo {year} {2019}{\natexlab{b}})}\BibitemShut {NoStop}%
\bibitem [{\citenamefont {N\'{e}vot}\ and\ \citenamefont
  {Croce}(1980)}]{nevot_caracterisation_1980}%
  \BibitemOpen
  \bibfield  {author} {\bibinfo {author} {\bibfnamefont {L.}~\bibnamefont
  {N\'{e}vot}}\ and\ \bibinfo {author} {\bibfnamefont {P.}~\bibnamefont
  {Croce}},\ }\href {\doibase 10.1051/rphysap:01980001503076100} {\bibfield
  {journal} {\bibinfo  {journal} {Rev. Phys. Appl. (Paris)}\ }\textbf {\bibinfo
  {volume} {15}},\ \bibinfo {pages} {761} (\bibinfo {year} {1980})}\BibitemShut
  {NoStop}%
\bibitem [{\citenamefont {Ku\v{c}erka}\ \emph {et~al.}(2004)\citenamefont
  {Ku\v{c}erka}, \citenamefont {Kiselev},\ and\ \citenamefont
  {Balgav\'{y}}}]{kucerka_determination_2004}%
  \BibitemOpen
  \bibfield  {author} {\bibinfo {author} {\bibfnamefont {N.}~\bibnamefont
  {Ku\v{c}erka}}, \bibinfo {author} {\bibfnamefont {M.~A.}\ \bibnamefont
  {Kiselev}}, \ and\ \bibinfo {author} {\bibfnamefont {P.}~\bibnamefont
  {Balgav\'{y}}},\ }\href {\doibase 10.1007/s00249-003-0349-0} {\bibfield
  {journal} {\bibinfo  {journal} {Eur. Biophys. J.}\ }\textbf {\bibinfo
  {volume} {33}},\ \bibinfo {pages} {328} (\bibinfo {year} {2004})}\BibitemShut
  {NoStop}%
\bibitem [{\citenamefont {Balgav\'{y}}\ \emph {et~al.}(2001)\citenamefont
  {Balgav\'{y}}, \citenamefont {Ku\v{c}erka}, \citenamefont {Gordeliy},\ and\
  \citenamefont {Cherezov}}]{balgavy_evaluation_2001}%
  \BibitemOpen
  \bibfield  {author} {\bibinfo {author} {\bibfnamefont {P.}~\bibnamefont
  {Balgav\'{y}}}, \bibinfo {author} {\bibfnamefont {N.}~\bibnamefont
  {Ku\v{c}erka}}, \bibinfo {author} {\bibfnamefont {V.~I.}\ \bibnamefont
  {Gordeliy}}, \ and\ \bibinfo {author} {\bibfnamefont {V.~G.}\ \bibnamefont
  {Cherezov}},\ }\href@noop {} {\bibfield  {journal} {\bibinfo  {journal}
  {Acta. Phys. Slovaca}\ }\textbf {\bibinfo {volume} {51}},\ \bibinfo {pages}
  {53} (\bibinfo {year} {2001})}\BibitemShut {NoStop}%
\bibitem [{\citenamefont {Tanford}(1980)}]{tanford_hydrophobic_1980}%
  \BibitemOpen
  \bibfield  {author} {\bibinfo {author} {\bibfnamefont {C.}~\bibnamefont
  {Tanford}},\ }\href@noop {} {\emph {\bibinfo {title} {The {{Hydrophobic
  Effect}}: {{Formation}} of {{Micelles}} and {{Biological Membranes}}}}},\
  \bibinfo {edition} {2nd}\ ed.\ (\bibinfo  {publisher} {{John Wiley \&
  Sons}},\ \bibinfo {address} {{New York, USA}},\ \bibinfo {year}
  {1980})\BibitemShut {NoStop}%
\bibitem [{\citenamefont {Braslau}\ \emph {et~al.}(1985)\citenamefont
  {Braslau}, \citenamefont {Deutsch}, \citenamefont {Pershan}, \citenamefont
  {Weiss}, \citenamefont {Als-Nielsen},\ and\ \citenamefont
  {Bohr}}]{braslau_surface_1985}%
  \BibitemOpen
  \bibfield  {author} {\bibinfo {author} {\bibfnamefont {A.}~\bibnamefont
  {Braslau}}, \bibinfo {author} {\bibfnamefont {M.}~\bibnamefont {Deutsch}},
  \bibinfo {author} {\bibfnamefont {P.~S.}\ \bibnamefont {Pershan}}, \bibinfo
  {author} {\bibfnamefont {A.~H.}\ \bibnamefont {Weiss}}, \bibinfo {author}
  {\bibfnamefont {J.}~\bibnamefont {Als-Nielsen}}, \ and\ \bibinfo {author}
  {\bibfnamefont {J.}~\bibnamefont {Bohr}},\ }\href {\doibase
  10.1103/PhysRevLett.54.114} {\bibfield  {journal} {\bibinfo  {journal} {Phys.
  Rev. Lett.}\ }\textbf {\bibinfo {volume} {54}},\ \bibinfo {pages} {114}
  (\bibinfo {year} {1985})}\BibitemShut {NoStop}%
\bibitem [{\citenamefont {Sinha}\ \emph {et~al.}(1988)\citenamefont {Sinha},
  \citenamefont {Sirota}, \citenamefont {Garoff},\ and\ \citenamefont
  {Stanley}}]{sinha_xray_1988}%
  \BibitemOpen
  \bibfield  {author} {\bibinfo {author} {\bibfnamefont {S.~K.}\ \bibnamefont
  {Sinha}}, \bibinfo {author} {\bibfnamefont {E.~B.}\ \bibnamefont {Sirota}},
  \bibinfo {author} {\bibfnamefont {S.}~\bibnamefont {Garoff}}, \ and\ \bibinfo
  {author} {\bibfnamefont {H.~B.}\ \bibnamefont {Stanley}},\ }\href {\doibase
  10.1103/PhysRevB.38.2297} {\bibfield  {journal} {\bibinfo  {journal} {Phys.
  Rev. B}\ }\textbf {\bibinfo {volume} {38}},\ \bibinfo {pages} {2297}
  (\bibinfo {year} {1988})}\BibitemShut {NoStop}%
\bibitem [{\citenamefont {Tikhonov}\ \emph {et~al.}(2000)\citenamefont
  {Tikhonov}, \citenamefont {Mitrinovic}, \citenamefont {Li}, \citenamefont
  {Huang},\ and\ \citenamefont {Schlossman}}]{tikhonov_xray_2000}%
  \BibitemOpen
  \bibfield  {author} {\bibinfo {author} {\bibfnamefont {A.~M.}\ \bibnamefont
  {Tikhonov}}, \bibinfo {author} {\bibfnamefont {D.~M.}\ \bibnamefont
  {Mitrinovic}}, \bibinfo {author} {\bibfnamefont {M.}~\bibnamefont {Li}},
  \bibinfo {author} {\bibfnamefont {Z.}~\bibnamefont {Huang}}, \ and\ \bibinfo
  {author} {\bibfnamefont {M.~L.}\ \bibnamefont {Schlossman}},\ }\href
  {\doibase 10.1021/jp001377u} {\bibfield  {journal} {\bibinfo  {journal} {J.
  Phys. Chem. B}\ }\textbf {\bibinfo {volume} {104}},\ \bibinfo {pages} {6336}
  (\bibinfo {year} {2000})}\BibitemShut {NoStop}%
\bibitem [{Note2()}]{Note2}%
  \BibitemOpen
  \bibinfo {note} {These values are tabulated in Table S1.}\BibitemShut {Stop}%
\bibitem [{\citenamefont {Li}\ \emph {et~al.}(1998)\citenamefont {Li},
  \citenamefont {Thirtle}, \citenamefont {Thomas}, \citenamefont {Penfold},
  \citenamefont {Webster},\ and\ \citenamefont {Rennie}}]{li_some_1998}%
  \BibitemOpen
  \bibfield  {author} {\bibinfo {author} {\bibfnamefont {Z.~X.}\ \bibnamefont
  {Li}}, \bibinfo {author} {\bibfnamefont {P.~N.}\ \bibnamefont {Thirtle}},
  \bibinfo {author} {\bibfnamefont {R.~K.}\ \bibnamefont {Thomas}}, \bibinfo
  {author} {\bibfnamefont {J.}~\bibnamefont {Penfold}}, \bibinfo {author}
  {\bibfnamefont {J.~R.~P.}\ \bibnamefont {Webster}}, \ and\ \bibinfo {author}
  {\bibfnamefont {A.~R.}\ \bibnamefont {Rennie}},\ }\href {\doibase
  10.1016/S0921-4526(98)00227-0} {\bibfield  {journal} {\bibinfo  {journal}
  {Physica B}\ }\textbf {\bibinfo {volume} {248}},\ \bibinfo {pages} {171}
  (\bibinfo {year} {1998})}\BibitemShut {NoStop}%
\bibitem [{\citenamefont {De{Wolf}}\ \emph {et~al.}(1999)\citenamefont
  {De{Wolf}}, \citenamefont {Leporatti}, \citenamefont {Kirsch}, \citenamefont
  {Klinger},\ and\ \citenamefont {Brezesinski}}]{dewolf_phase_1999}%
  \BibitemOpen
  \bibfield  {author} {\bibinfo {author} {\bibfnamefont {C.}~\bibnamefont
  {De{Wolf}}}, \bibinfo {author} {\bibfnamefont {S.}~\bibnamefont {Leporatti}},
  \bibinfo {author} {\bibfnamefont {C.}~\bibnamefont {Kirsch}}, \bibinfo
  {author} {\bibfnamefont {R.}~\bibnamefont {Klinger}}, \ and\ \bibinfo
  {author} {\bibfnamefont {G.}~\bibnamefont {Brezesinski}},\ }\href {\doibase
  10.1016/S0009-3084(98)00104-2} {\bibfield  {journal} {\bibinfo  {journal}
  {Chem. Phys. Lipids}\ }\textbf {\bibinfo {volume} {97}},\ \bibinfo {pages}
  {129} (\bibinfo {year} {1999})}\BibitemShut {NoStop}%
\bibitem [{\citenamefont {Brezesinski}\ \emph {et~al.}(2001)\citenamefont
  {Brezesinski}, \citenamefont {M\"{u}ller}, \citenamefont {Toca-Herrera},\
  and\ \citenamefont {Krustev}}]{brezesinski_xray_2001}%
  \BibitemOpen
  \bibfield  {author} {\bibinfo {author} {\bibfnamefont {G.}~\bibnamefont
  {Brezesinski}}, \bibinfo {author} {\bibfnamefont {H.~J.}\ \bibnamefont
  {M\"{u}ller}}, \bibinfo {author} {\bibfnamefont {J.~L.}\ \bibnamefont
  {Toca-Herrera}}, \ and\ \bibinfo {author} {\bibfnamefont {R.}~\bibnamefont
  {Krustev}},\ }\href {\doibase 10.1016/S0009-3084(01)00135-9} {\bibfield
  {journal} {\bibinfo  {journal} {Chem. Phys. Lipids}\ }\textbf {\bibinfo
  {volume} {110}},\ \bibinfo {pages} {183} (\bibinfo {year}
  {2001})}\BibitemShut {NoStop}%
\bibitem [{\citenamefont {Struth}\ \emph {et~al.}(2001)\citenamefont {Struth},
  \citenamefont {Rieutord}, \citenamefont {Konovalov}, \citenamefont
  {Brezesinski}, \citenamefont {Gr\"{u}bel},\ and\ \citenamefont
  {Terech}}]{struth_organization_2001}%
  \BibitemOpen
  \bibfield  {author} {\bibinfo {author} {\bibfnamefont {B.}~\bibnamefont
  {Struth}}, \bibinfo {author} {\bibfnamefont {F.}~\bibnamefont {Rieutord}},
  \bibinfo {author} {\bibfnamefont {O.}~\bibnamefont {Konovalov}}, \bibinfo
  {author} {\bibfnamefont {G.}~\bibnamefont {Brezesinski}}, \bibinfo {author}
  {\bibfnamefont {G.}~\bibnamefont {Gr\"{u}bel}}, \ and\ \bibinfo {author}
  {\bibfnamefont {P.}~\bibnamefont {Terech}},\ }\href {\doibase
  10.1103/PhysRevLett.88.025502} {\bibfield  {journal} {\bibinfo  {journal}
  {Phys. Rev. Lett.}\ }\textbf {\bibinfo {volume} {88}},\ \bibinfo {pages}
  {025502} (\bibinfo {year} {2001})}\BibitemShut {NoStop}%
\bibitem [{\citenamefont {Sivia}\ \emph {et~al.}(1993)\citenamefont {Sivia},
  \citenamefont {David}, \citenamefont {Knight},\ and\ \citenamefont
  {Gull}}]{sivia_introduction_1993}%
  \BibitemOpen
  \bibfield  {author} {\bibinfo {author} {\bibfnamefont {D.~S.}\ \bibnamefont
  {Sivia}}, \bibinfo {author} {\bibfnamefont {W.~I.~F.}\ \bibnamefont {David}},
  \bibinfo {author} {\bibfnamefont {K.~S.}\ \bibnamefont {Knight}}, \ and\
  \bibinfo {author} {\bibfnamefont {S.~F.}\ \bibnamefont {Gull}},\ }\href
  {\doibase 10.1016/0167-2789(93)90241-R} {\bibfield  {journal} {\bibinfo
  {journal} {Physica D}\ }\textbf {\bibinfo {volume} {66}},\ \bibinfo {pages}
  {234} (\bibinfo {year} {1993})}\BibitemShut {NoStop}%
\end{thebibliography}%
\end{document}